\documentclass[epj]{svjour}
\usepackage{graphics}
\usepackage{amsmath,amssymb,color}

\begin{document}

\title{Urban road networks -- Spatial networks with universal geometric features?} 
\subtitle{A case study on Germany's largest cities}
\titlerunning{Urban road networks -- Spatial networks with universal geometric features?}

\author{Sonic H.~Y. Chan\inst{1,2}\and Reik V. Donner\inst{1,3}\and Stefan L\"ammer\inst{1}}
\institute{Institute for Transport and Economics, Dresden University of Technology, W\"urzburger Str.~35, 01187 Dresden, Germany
\and {Department of Physics, The Chinese University of Hong Kong, Shatin, N.T., Hong Kong}
\and Potsdam Institute for Climate Impact Research, P.O. Box 60\,12\,03, 14412 Potsdam, Germany}
\mail{\\Reik V. Donner (\texttt{reik.donner@pik-potsdam.de})}
\date{Received: \today / Revised version: }

\abstract{Urban road networks have distinct geometric properties that are partially determined by their (quasi-) two-dimensional structure. In this work, we study these properties for 20 of the largest German cities. We find that the small-scale geometry of all examined road networks is extremely similar. The object-size distributions of road segments and the resulting cellular structures are characterised by heavy tails. As a specific feature, a large degree of rectangularity is observed in all networks, with link angle distributions approximately described by stretched exponential functions. We present a rigorous statistical analysis of the main geometric characteristics and discuss their mutual interrelationships. Our results demonstrate the fundamental importance of cost-efficiency constraints for in time evolution of urban road networks.}

\PACS{{89.40.Bb}{Land transportation} \and {89.75.Fb}{Structures and organisation in complex systems} \and {89.75.Hc}{Networks and genealogical trees}}

\maketitle

\section{Introduction}

During the last decade, the understanding of the structure and dynamics of complex networks and their mutual interplay has made enormous progress \cite{Strogatz2001,Albert2002,Dorogovtsev2002,Newman2003,Boccaletti2006,Watts1999,Bornholdt2003,Dorogovtsev2003,Pastor-Satorras2003,Ben-Naim2004}. A specific type of networks are transportation systems, in which the directed flow of individual transportation units is often bounded to a spatially discrete substrate consisting of mutually connected tracks. The merges, diverges, and crossings between these tracks can be considered as the network nodes, and their physical connections as the corresponding links. The resulting graph properties determine -- together with the link capacities and the actual spatio-temporal demand pattern -- the efficiency of transport in the system.

In modern society, the optimal performance of transportation systems is of paramount importance for daily life. As a consequence, the structure and time evolution of such networks often result from careful planning processes. Numerous characteristic parameters have been studied for rail- and subway \cite{Marchiori2000,Latora2001,Latora2002,Sen2003,Seaton2004,Vragovic2005,Kurant2006a,Kurant2006b,Chang2006,Xu2007,Li2007,Lee2008,Ru2008,Domenech2009}, airport \cite{Amaral2000,Chi2003,Li2004,Barrat2004,Guimera2004,Guimera2005,Wang2005,Li2006b,Guida2007,Bagler2008,Bagler2009,Correa2008}, urban transit \cite{Wu2004}, public transportation \cite{Sienkiewicz2005,Sienkiewicz2005b,Gastner2006b,Li2006,vonFerber2007,vonFerber2009,Berche2009}, and maritime transport networks \cite{Xu2007b,Hu2008,Kaluza2010} on a variety of different spatial scales, revealing that many characteristics do strongly differ between different systems. Amaral \textit{et~al.}~\cite{Amaral2000} used this observation for a classification of small-world networks (i.e., graphs with a short average path length and a high degree of clustering) into scale-free, broad-scale and single-scale networks. For example, the world airport network has been classified as a single-scale network \cite{Amaral2000}, whereas the node degree distributions of the Italian airport network \cite{Guida2007} and the Chinese railway network \cite{Li2007} are scale-free. Gastner and Newman \cite{Gastner2006} proposed a connection between the formation of airport and highway networks. Although both types of networks are intrinsically different, their statistical features can be understood as the result of the same optimisation problem with one parameter in the goal function varied, leading to a variety of network topologies. 

In contrast to many networks in other areas of research (e.g., biology, telecommunication, or social sciences), urban road systems can be (in good approximation) considered as \textit{planar} networks, i.e., links cannot ``cross'' each other without forming a physical intersection (node) as long as there are no tunnels or bridges. Therefore, the study of urban road networks allows distinguishing which statistical properties are actually specific for planar transportation systems. {Due to reasons of practicality, which will be further studied in the course of this paper, and given} the planarity constraint, it follows that the node degrees are {practically} limited (i.e., the networks cannot be scale-free) and the average path length is much larger than for scale-free networks with a comparable number of nodes. Both features are therefore distinctively different from those of most elementary evolving network models, such as (Erd\H{o}s-R\'enyi) random graphs \cite{Erdoes1959,Bollobas2001}, (B\'a\-ra\-ba\-si-Albert) scale-free \cite{Barabasi1999,Dorogovtsev2000,Caldarelli2007}, or (Watts-Strogatz) small-world networks \cite{Watts1999}. Even more, the historical development of cities can often be understood as the result of certain self-organisation processes, i.e., road networks usually have not been globally designed following some general growth mechanisms of urban areas \cite{Schaur1991,Frankhauser1994,Schweitzer1997,Weidlich1999,Weidlich2000,Humpert2002,Schweitzer2003,Batty2005}, which are characterised by scaling laws in various urban supply networks and related growth processes \cite{Kuehnert2006,Bettencourt2007,Helbing2009}. 

\begin{figure*}[thbp]
\begin{center}
\resizebox{\columnwidth}{!}{\includegraphics{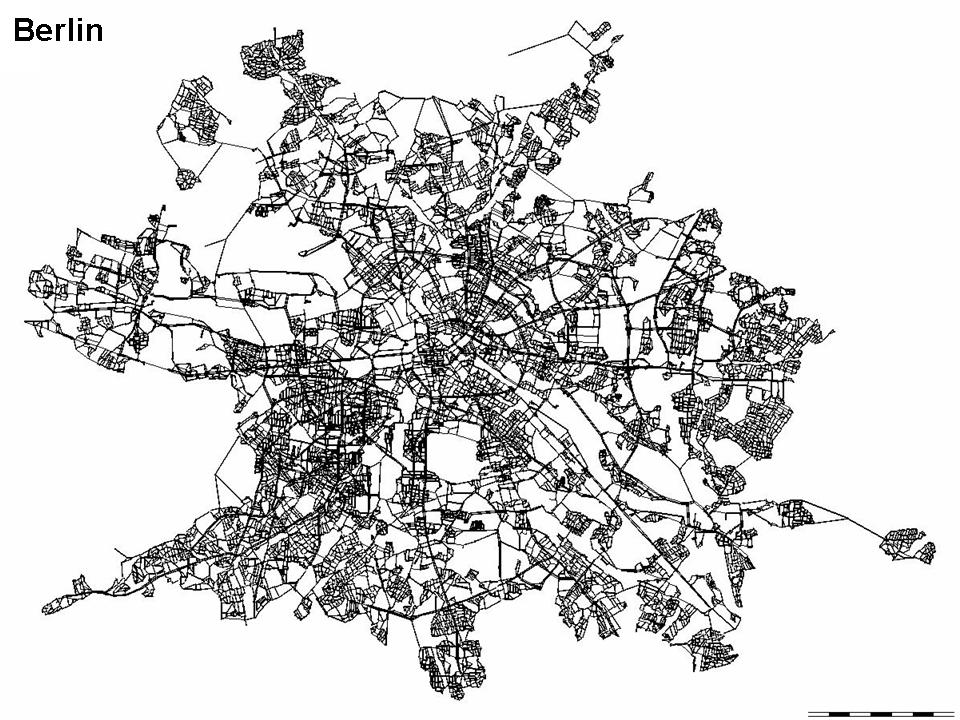}} \hfill
\resizebox{\columnwidth}{!}{\includegraphics{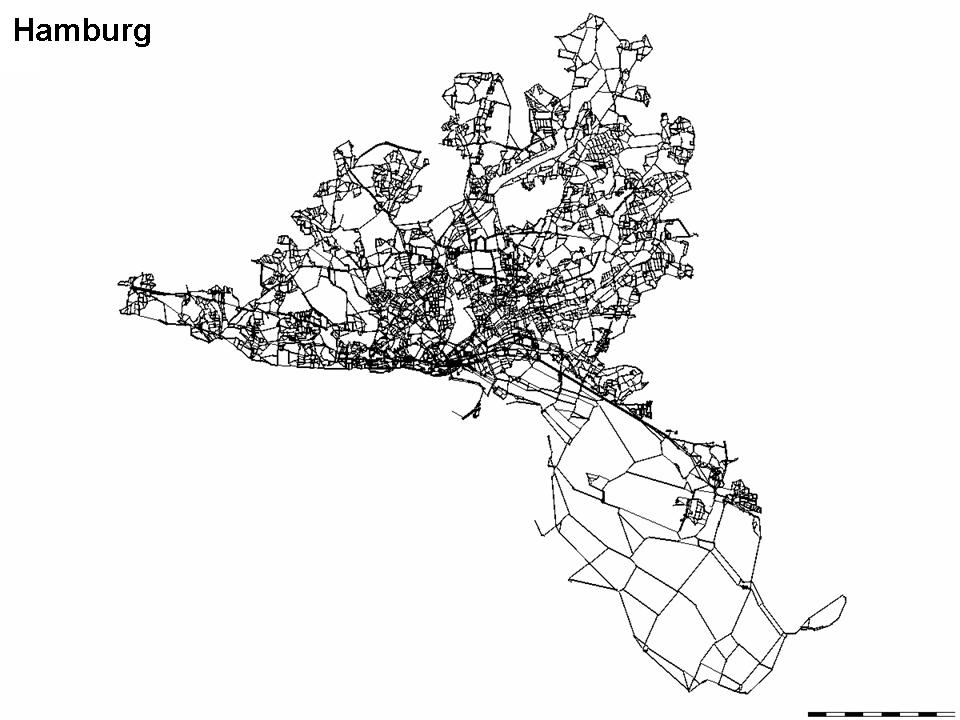}} \\
\resizebox{\columnwidth}{!}{\includegraphics{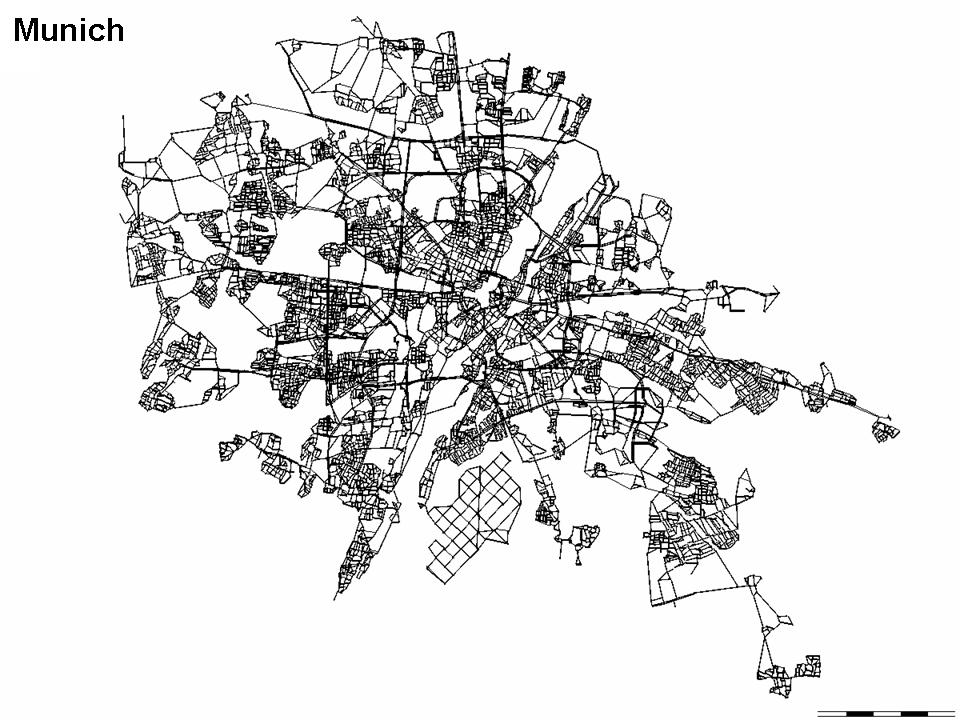}} \hfill
\resizebox{\columnwidth}{!}{\includegraphics{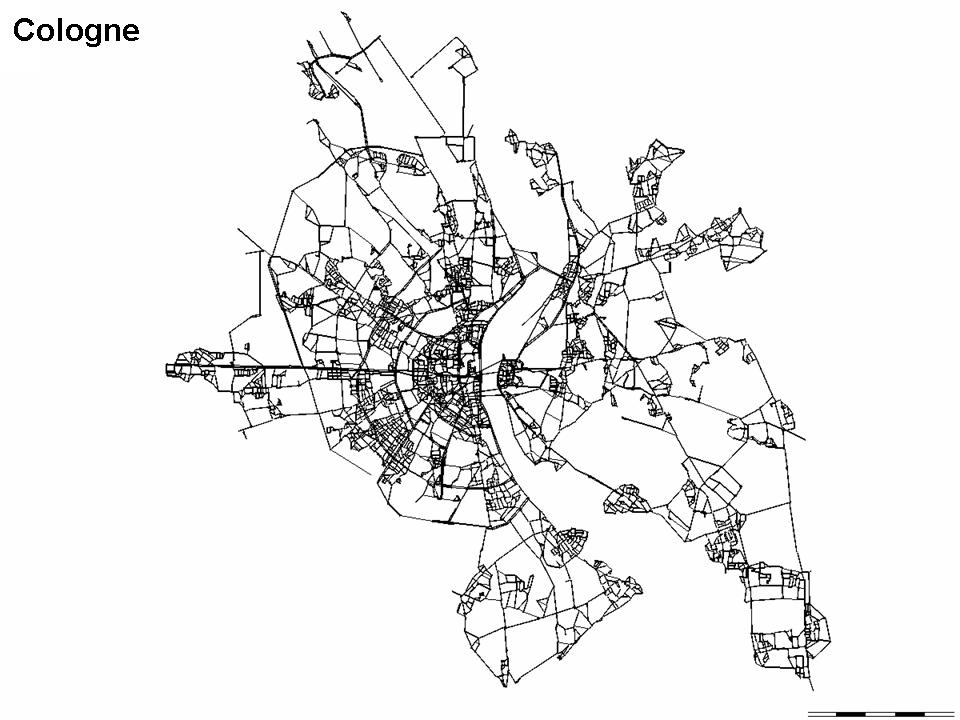}} 	
\end{center}
\caption{Large-scale appearance of the road networks of the four largest German cities. The respective scales indicate basic units of 1 km length.}
\label{fig:example}
\end{figure*}

While many results obtained so far considered the properties of the dual (or connectivity) graphs associated with urban road networks (i.e., graphs where roads are represented by nodes, which are linked if the physical roads do ever mutually intersect \cite{Jiang2004a,Jiang2004b}), in this contribution, we will study and thoroughly compare the characteristic \textit{geometric} properties on the level of the elementary units (intersections, road segments, cellular structures) in \textit{physical} space \cite{Laemmer2006a}. Unlike in past related studies focussing on rather small road networks, the aim of this work is to better understand the specific geometric features of the \textit{complete} urban road networks of present-day cities, disregaring their specific historical development. In order to study a homogeneous sample, we consider 20 of the largest German cities, which are characterised by rather different traffic conditions and large-scale morphologies (see Fig.~\ref{fig:example} for some examples). In this paper, we will show, however, that the small-scale geometric properties of the corresponding road networks show a very large degree of similarity, suggesting that the common features determining the network topology are the planarity and the requirement of an efficient use of the available space, which leads to the preferred emergence of rectangular structures. 

This paper is organised as follows: In Section 2, the study area and used data are described. Sections 3 to 5 summarise our results on the statistical properties of nodes, links, and cells as the elementary geometric units of a road network. A detailed discussion of our findings in relationship to recent results from other studies is presented in Section 6. Finally, some outlook on further relevant directions of research is given.

\section{Study Area}
\label{sec:Structure}

In this work, we thoroughly extend a recent study \cite{Laemmer2006a} on the urban road networks of 20 of the largest German cities. The \textit{large-scale} morphology of these networks, which is determined by geographical restrictions (hillslope, rivers, soil properties, etc.) and the particular historical development (creation of rail- or highways, emergence of certain functional requirements, war damages, etc.), differs considerably between the individual cities (Fig.~\ref{fig:example}).

Our data have been obtained for the year 2005 from the geographic data base TeleAtlas MultiNet\texttrademark (for detailed information, see \texttt{www.teleatlas.com}), which is used in numerous applications such as Google\texttrademark Maps, real-time navigation systems (TomTom), and urban planning and management. From the 21 largest cities, only Hannover has not been taken into account because of a significantly lower spatial resolution of the urban area in the initial data base, which can be seen from the spatial density of intersections (nodes) being by a factor of 5-20 smaller than for the other 20 top-ranked cities (see Tab.~\ref{tab:GermanChar12}). Moreover, since the initial data base has not been sufficiently homogeneous (e.g., in terms of the representation of separate lanes), we have applied a two-step preprocessing: 

\begin{table*}[thb]
\renewcommand{\arraystretch}{0.7}
\tabcolsep=6pt
	\begin{tabular}{llrrrrrrrrrr}
		\hline
		Rank & \multicolumn{1}{c}{City} & \multicolumn{1}{c}{Population} & \multicolumn{1}{c}{Area} & \multicolumn{4}{c}{Original} & \multicolumn{3}{c}{Processed} & \multicolumn{1}{c}{$\rho_n$} \\
		 &  &  & (km$^2$) & $N_n$ & $N_l^*$ & $N_l$ & $N_c$ & $N_n$ & $N_l$ & $N_c$ & [km$^{-2}$] \\
		\hline
		1 & Berlin & 3,392,425 & 891 & 37,020 & 87,795 & 50,632 & 13,545 & 19,931 & 33,000 & 12,929 & 22.4 \\
		2 & Hamburg & 1,728,806 & 753 & 19,717 & 43,819 & 25,641 & 5,867 & 9,044 & 14,603 & 5,458 & 12.0 \\
		3 & Munich & 1,234,692 & 311 & 21,393 & 49,521 & 28,579 & 7,074 & 11,058 & 17,934 & 6,737 & 35.6 \\
		4 & Cologne & 968,639 & 405 & 14,553 & 29,359 & 18,182 & 3,602 & 5,395 & 8,724 & 3,259 & 13.3 \\
		5 & Frankfurt & 643,726 & 249 & 9,728 & 18,104 & 12,366 & 2,622 & 3,911 & 6,394 & 2,445 & 15.7 \\
		6 & Dortmund & 590,831 & 281 & 10,326 & 22,579 & 12,512 & 2,176 & 3,281 & 5,249 & 1,923 & 11.7 \\
		7 & Stuttgart & 588,477 & 208 & 10,302 & 21,934 & 12,748 & 2,414 & 3,612 & 5,848 & 2,164 & 17.4 \\
		8 & Essen & 585,481 & 210 & 11,387 & 24,537 & 14,124 & 2,704 & 4,093 & 6,598 & 2,444 & 19.5 \\
		9 & Dusseldorf & 571,886 & 218 & 8,237 & 16,773 & 10,377 & 2,114 & 3,124 & 5,099 & 1,922 & 14.3 \\
		10 & Bremen & 542,987 & 318 & 10,227 & 21,702 & 12,748 & 2,477 & 3,827 & 6,106 & 2,198 & 12.0 \\
		\it 11 & \it Hannover & \it 517,310 & \it 204 & \it 1,589 & \it 3,463 & \it 1,856 & \it 266 & \it 411 & \it 640 & \it 225 & \it 2.0 \rm \\
		12 & Duisburg & 508,664 & 233 & 6,300 & 14,333 & 8,141 & 1,832 & 2,837 & 4,573 & 1,710 & 12.2 \\
		13 & Leipzig & 494,795 & 293 & 9,071 & 21,199 & 11,643 & 2,561 & 3,753 & 6,137 & 2,355 & 12.8 \\
		14 & Nuremberg & 493,397 & 187 & 8,768 & 18,639 & 11,157 & 2,339 & 3,543 & 5,762 & 2,141 & 18.9 \\
		15 & Dresden & 480,228 & 328 & 9,643 & 22,307 & 11,977 & 2,321 & 3,346 & 5,488 & 2,109 & 10.2 \\
		16 & Bochum & 388,869 & 146 & 6,970 & 15,091 & 8,437 & 1,455 & 2,233 & 3,555 & 1,276 & 15.3 \\
		17 & Wuppertal & 363,522 & 168 & 5,681 & 11,847 & 6,869 & 1,177 & 1,750 & 2,803 & 1,029 & 10.4 \\
		18 & Bielefeld & 324,815 & 259 & 8,259 & 18,280 & 9,880 & 1,613 & 2,546 & 4,035 & 1,461 & 9.8 \\
		19 & Bonn & 308,921 & 141 & 6,365 & 13,746 & 7,762 & 1,381 & 2,094 & 3,365 & 1,238 & 14.9 \\
		20 & Mannheim & 308,759 & 145 & 5,819 & 12,581 & 7,607 & 1,765 & 2,674 & 4,364 & 1,652 & 18.4 \\
		21 & Karlsruhe & 281,334 & 173 & 5,505 & 11,866 & 7,004 & 1,488 & 2,204 & 3,594 & 1,366 & 12.7 \\
		\hline
	\end{tabular}
	\caption{Characteristics of the urban road networks of the 21 largest German cities in the original and the processed data: number of nodes $N_n$, links $N_l$, and cells $N_c$ in the network. The quantity $N_l^*$ corresponds to the content of the original data base with partially directed links, whereas $N_l$ reflects the number of links in the reduced undirected network data base. In addition, the spatial density $\rho_n$ of intersections in the processed data is given.}
	\label{tab:GermanChar12}
\end{table*}

In a first stage, multiple and partially directed links have been combined by defining that each pair of nodes can only be connected by at most one undirected link. Removing replications and directional information, the number of links forming the network (and, hence, the average degree of the nodes) is substantially reduced (see Tab.\ \ref{tab:GermanChar12}). In the following, the resulting undirected network representation will be referred to as the \textit{original data}. 

In order to better understand some of the geometric features of the networks, a second preprocessing step has been applied. First, all nodes of degree $k_n=2$ that represent curved or sharply bended roads rather than physical intersections of different streets have been removed from the data base. This leads to an additional decrease in the number of both nodes and links in the networks. Second, for convenience, all dead-end roads have been removed from the network, since such roads also have only minor effects on the traffic conditions in the networks. Formally, this has been realised by excluding all nodes of degree $k_n=1$ and their connecting links from the original data. Wherever this procedure reduced the degree of other nodes to $k_n=2$, these have also been removed.

\section{Node Properties}\label{sec:nodes}

Recent studies on urban road networks have often focussed on the statistical properties of the nodes. Apart from their spatial density (which is additionally reflected in the link and cell properties, which will be discussed in Sections \ref{sec:links} and \ref{sec:cells}, respectively), nodes are mainly characterised by their degree distribution $p(k_n)$, where the degree $k_n$ of a node in an undirected network is defined as the number of road segments (links) that are connected at the respective position in geographical space. 

Several authors have already considered the degree distribution in different road networks. Buhl \textit{et~al.}~\cite{Buhl2006} studied 41 road systems from small- to intermediate-sized cities around the world. For the average node degree $\left<k_n\right>$, they found values between 2.02 and 2.86. The observed frequency distributions of the node degrees showed an exponential tail for $k_n\geq 3$ with a variety of possible characteristic scale parameters, indicating the presence of single-scale networks. Scellato \textit{et~al.}~\cite{Scellato2006} compared the properties of the road network of the historically grown city of Bologna with the planned square-like grid structure of San Francisco, yielding average node degrees of $\left<k_n\right>=2.71$ and $3.21$, respectively. Cardillo \textit{et~al.}~\cite{Cardillo2006} argued that this observation can be explained by the extraordinarily high abundance of ``full'' intersections ($k_n=4$) in grid-like cities, with $p(k_n=4)>p(k_n=3)$, whereas $k_n=3$ is much more abundant in naturally grown settlements. Crucitti and co-workers \cite{Crucitti2006a,Crucitti2006b,Porta2006a,Porta2006b} demonstrated that self-organised and planned cities also differ in the probability distributions $p(l_l)$ of the link lengths (unimodal for naturally grown, but irregularly shaped for planned cities) and the scaling behaviour of different centrality measures.

\begin{table}[htb]
\renewcommand{\arraystretch}{0.7}
\tabcolsep=6pt
	\begin{tabular}{r r r@{$\times$}l c}
		\hline
		$k_n$ & \multicolumn{1}{c}{$N(k_n)$} & \multicolumn{2}{c}{$p(k_n)$} & \multicolumn{1}{c}{$\left<l_l\right>$ [m] ($\sigma_{l_l}$ [m])}\\
		\hline
		3 & 71,232 & 7.52 & $10^{-1}$ & 157.63 (202.01) \\
		4 & 23,150 & 2.44 & $10^{-1}$ & 142.91 (164.54) \\
		5 & 265 & 2.80 & $10^{-3}$ & 125.32 (111.81) \\
		6 & 19 & 2.01 & $10^{-4}$ & 126.04 (170.06) \\
		7 & 1 & 1.06 & $10^{-5}$ & 112.36 (104.04) \\
		\hline
	\end{tabular}
	\caption{Absolute and relative frequencies of node degrees $k_n$ in the urban road networks of German cities, and dependence of the mean lengths $\left<l_l\right>$ and associated standard deviations of the connected links on the node degrees.}
	\label{tab:German_nkP}
\end{table}

In the German road networks, we observe that nodes with $k_n=4$ occur much more often than one would expect for a single-scale network (i.e., a network with an exponentially decaying degree distribution, see Tab.\ \ref{tab:German_nkP}), which is consistent with previous studies. This very high abundance obviously results from intersections between continuing roads. Although the tail of the degree distribution resembles an exponential decay, due to the low maximum node degree in planar networks, it is hardly possible to statistically identify any particular type of scaling \cite{Kalapala2006}. 

\begin{table*}[thb]
\renewcommand{\arraystretch}{0.7}
\tabcolsep=6pt
	\begin{tabular}{llccccccrc}
		\hline
		Rank & City & $\left<k_n\right>$ & $\left<l_l\right>$/[m] & $\left<\theta_l\right>$/[$^o$] & 
		$\left<\cos^2 2\theta_l\right>$ & $\left<\mu\right>$ &
		$\left<k_c\right>$ & $\left<A_c\right>$/[m$^2$] & $\left<\phi_c\right>$ \\
		& & & & & $(\sigma_{\cos^2 2\theta_l})$ & $(\sigma_{\mu})$ & & & $(\sigma_{\phi_c})$ \\
		\hline
		
1	& Berlin & 3.311 & 149.65 &	108.604 &	0.757 (0.322) & 0.317 (0.371) & 4.999 & 42,464.47 &	0.385 (0.166) \\
2	& Hamburg	& 3.229 &	166.01 & 111.721 & 0.696 (0.335) & 0.402 (0.411) & 5.199 & 67,598.39 & 0.365 (0.166) \\
3	& Munich & 3.244 & 140.07 &	110.978 &	0.768 (0.310) & 0.312 (0.365) & 5.147 &	36,109.65 &	0.409 (0.159) \\
4	& Cologne &	3.234 &	145.20 & 111.670 & 0.712 (0.337) & 0.371 (0.392) & 5.127 & 50,105.39 & 0.388 (0.167) \\
5	& Frankfurt	& 3.270	& 135.64 & 110.361 & 0.721 (0.336) & 0.358 (0.386) & 5.017 & 35,104.61 & 0.394 (0.169) \\
6	& Dortmund & 3.200 & 182.39	& 112.926 &	0.698 (0.341) & 0.393 (0.406) & 5.273 &	86,160.23 &	0.385 (0.168) \\
7	& Stuttgart	& 3.238	& 140.78 & 110.929 & 0.697 (0.345) & 0.406 (0.431) & 5.031 & 36,760.97 & 0.371 (0.173) \\
8	& Essen	& 3.224	& 148.29 & 112.288 & 0.672 (0.344) & 0.419 (0.406) & 5.178 & 41,400.13 & 0.378 (0.169) \\
9	& Dusseldorf & 3.264 & 150.57	& 111.007 &	0.698 (0.338) & 0.396 (0.407) & 5.007 &	39,061.47 &	0.367 (0.175) \\
10 & Bremen	& 3.191	& 164.42 & 113.145 & 0.729 (0.327) & 0.371 (0.414) & 5.150 & 47,000.51 & 0.352 (0.164) \\
\it 11 & \it Hannover	& \it 3.114	& \it 193.55 & \it 114.973 & \it 0.635 (0.355) & \it 0.453 (0.408) & \it 5.102 & \it 63,994.27 & \it 0.387 (0.151) \rm \\
12 & Duisburg	& 3.224	& 142.97 & 111.376 & 0.735 (0.328) & 0.356 (0.400) & 5.047 & 42,777.24 & 0.401 (0.171) \\
13 & Leipzig & 3.270 & 155.85	& 109.987 &	0.739 (0.224) & 0.336 (0.385) & 5.036 &	52,764.49 &	0.407 (0.160) \\
14 & Nuremberg & 3.253 & 147.49	& 110.871 &	0.716 (0.330) & 0.380 (0.403) & 5.046 &	42,845.17 &	0.381 (0.173) \\
15 & Dresden & 3.280 & 177.39 &	110.281 &	0.712 (0.342) & 0.365 (0.391) & 5.015 &	66,686.65 &	0.411 (0.159) \\
16 & Bochum	& 3.184	& 186.16 & 113.300 & 0.660 (0.349) & 0.439 (0.420) & 5.258 & 70,223.26 & 0.370 (0.168) \\
17 & Wuppertal & 3.203 & 147.35	& 112.859 &	0.658 (0.352) & 0.439 (0.422) & 5.185 &	43,051.27 &	0.369 (0.163) \\
18 & Bielefeld & 3.170 & 199.27	& 114.057 &	0.691 (0.336) & 0.408 (0.411) & 5.338 &	100,863.19 &	0.384 (0.161) \\
19 & Bonn	& 3.214	& 153.29 & 112.504 & 0.702 (0.337) & 0.394 (0.408) & 5.060 & 38,561.61 & 0.383 (0.172) \\
20 & Mannheim	& 3.264	& 128.32 & 110.336 & 0.763 (0.330) & 0.313 (0.388) & 4.999 & 32,144.16 & 0.411 (0.162) \\
21 & Karlsruhe & 3.261 & 145.42	& 110.663 &	0.751 (0.318) & 0.343 (0.398) & 4.947 &	32,447.63 &	0.377 (0.167) \\
  \hline
  & All cities & 3.251 & 153.04 & 111.034 & 0.726 (0.332) & 0.361 (0.395) & 5.090 & 48,494.07 & 0.386 (0.167) \\
  \hline
	\end{tabular}
	\caption{Geometric characteristics of the urban road networks of the 21 largest German cities: node degree $k_n$, link length $l_l$, link angle $\theta_l$, rectangularity parameters $\cos^2 2\theta_l$ and $\mu$, topological cell degree $k_c$, cell area $A_c$, and form factor $\phi_c$. $\left<\cdot\right>$ and $\sigma_{\cdot}$ represent mean values and standard deviations taken over all objects in the respective city. Note that the results partially differ from those in \cite{Laemmer2006a} due to the different preprocessing of the data in {our} study.}
	\label{tab:GermanChar12b}
\end{table*}

We note that the average node degrees $\left<k_n\right>$ of the individual cities lie within a very narrow range between 3.17 and 3.31 (see Tab.~\ref{tab:GermanChar12b}), although the spatial density of nodes differs strongly (Tab.~\ref{tab:GermanChar12}). The average value for all cities, $\left<k_n\right>=3.25$, is remarkably higher than that reported for the US interstate highway network ($\left<k_n\right>=2.86$ \cite{Gastner2006}), but still far below the theoretical limit for planar networks ($\left<k_n\right>\leq 6$ \cite{West1996}). The fact that urban road networks have a higher mean node degree than highway networks is primarily related to the much higher abundance of nodes of degree $k_n=4$. As it will be discussed later, the abundance of nodes with degrees 3 and 4 is closely related to the dominating presence of perpendicular structures in urban road networks, which is most probably determined by both architectural and economic requirements. 

Comparing the results for the German cities with the outcomes of other recent studies, it is remarkable that both Buhl \textit{et~al.} \cite{Buhl2006} and Scellato \textit{et~al.} \cite{Scellato2006} found lower values of the mean degree, which can however be easily explained by our preprocessing removing all nodes of degree $k_n=1$ and 2. If these nodes had been included in the data, the mean node degrees would have taken values between 2.393 (Bielefeld) and 2.735 (Berlin), which is in reasonable agreement with the results of \cite{Buhl2006,Scellato2006}. The latter values follow directly from Tab.~\ref{tab:GermanChar12}, since for undirected planar networks, the mean node degree is determined by
\begin{equation}
\left<k_n\right>=2\frac{N_l}{N_n},
\end{equation}
\noindent
where $N_l$ and $N_n$ are the total numbers of links and nodes, respectively.

The fast decay of the node degree distribution is a direct consequence of the planarity of road networks. In terms of practical usage, road intersections having too many connections do not provide a feasible solution for controlling conflicting traffic flows. Instead, at such intersections the total waiting times are significantly enhanced and bottlenecks are formed, reducing the overall transportation efficiency on the network.

\section{Link Properties} \label{sec:links}

The link-based geometric properties of a road network are closely related to the spatial distribution and to the degree of the nodes. However, we note that links do not always connect neighbouring nodes along their shortest distance, which is due to secondary factors like preferred building geometry, soil properties, or hillslope. Thus, one has to distinguish between the properties of the physical (curved) links and those of the Euclidean (shortest-distance) distances between their starting and end points. In the latter case, all existing links are assumed to be non-curved (cf. our preprocessing), resulting in a systematic bias towards shorter link lengths. In a similar spirit, depending on whether the original or processed data are used, the term link itself can be identified with either a straight road segment without any curves and interior connections to other roads (i.e., nodes with $k_n=2$ exist) or the actual physical (possibly curved) connection between two practically relevant nodes ($k_n\geq 3$). In the first case, the links are systematically shorter than in the second one. 

Unless it is explicitly stated otherwise, in the following, we will consider the properties of the ``physical'', potentially curved links. For this purpose, we will define the link lengths and directions according to the original data, but combining several links into one whenever intermediate nodes have been removed in our second preprocessing step (i.e., respective node positions and degrees are taken from the processed data).

\subsection{Link length distributions}

Travel time is probably the most important indicator for the efficiency of a road network. However, this quantity strongly depends on secondary parameters such as the speed limit and the particular traffic situation (involving factors like time of the day, spatial distribution of different functional areas, etc.). Because of this, we consider here the travel distance rather than travel time, which is measured by the physical length $l_l$ of each link. 

{Let us first discuss the mean link lengths for different cities.} Table~\ref{tab:GermanChar12b} shows that {this property varies} remarkably (between 128 m (Mannheim) and 199 m (Bielefeld)) {among the different cities}. In particular, there is a tendency that cities with a small average node degree are constructed by somewhat longer road segments. Quantifying this trend by the linear Pearson correlation coefficient $r$ and Spearman's rank-order correlation coefficient $\rho$ as a related dis\-tri\-bu\-tion-free measure yields values of $r=-0.55$ and $\rho=-0.42$, respectively, which supports this finding, but underlines that the variations of the mean link lengths are not exclusively determined by the average node degrees.

\begin{figure}[t!]
\begin{center}
\resizebox{0.48\columnwidth}{!}{\includegraphics{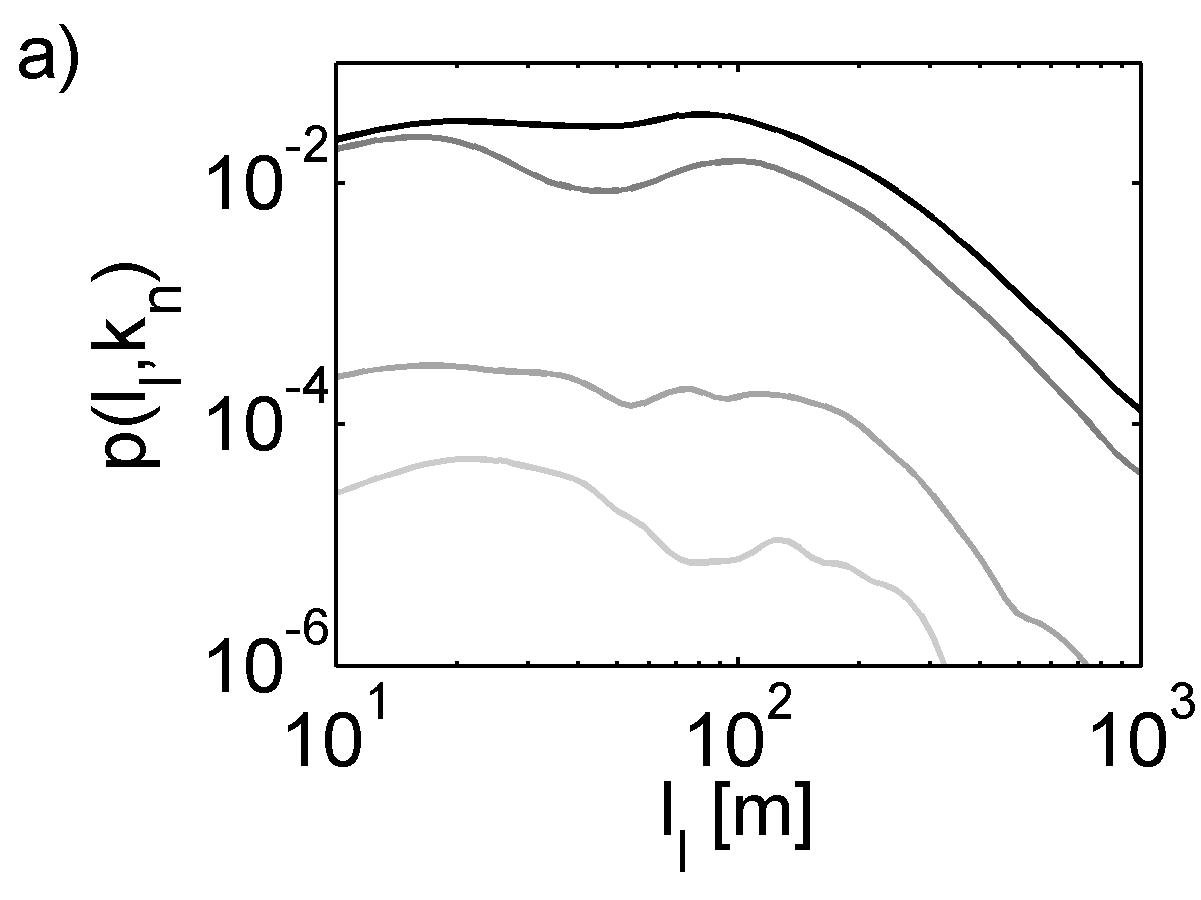}} \hfill
\resizebox{0.48\columnwidth}{!}{\includegraphics{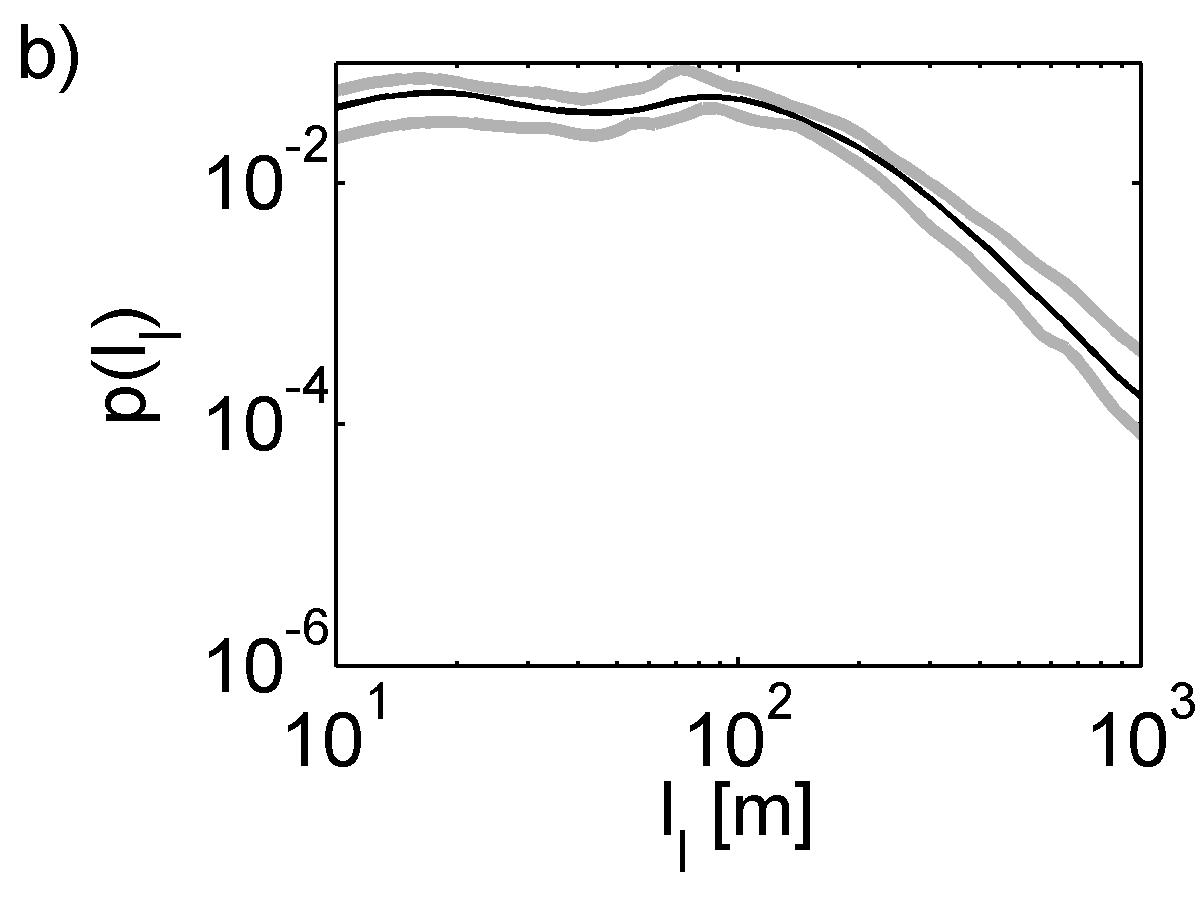}} 
\end{center}
\caption{a) Relative frequency distribution $p(l_l,k_n)=p(l_l|k_n)p(k_n)$ of the link lengths of all cities for node degrees of $k_n=3$, 4, 5, and 6 (from top to bottom and black to light grey). b) Relative frequency distribution $p(l_l)$ of the link lengths for all 20 cities (black), and upper and lower limits obtained from the individual cities (grey lines).}
\label{fig:ll1}
\end{figure}

{In order to get a more complete picture, we furthermore study the full frequency distributions of link lengths in the different networks. As a first notable observation regarding the local interrelationships between link and node properties,} Tab.~\ref{tab:German_nkP} {demonstrates that} the link lengths decrease {on average} with increasing degree of the involved nodes. {Considering the link length distributions explicitly} in dependence on $k_n$, Fig.~\ref{fig:ll1} {reveals that} the qualitative behaviour is similar for all node degrees and consists of two disjoint regimes: (i) a saturation in the abundance of link lengths below 100 m, which seems to result from a {weakly} bimodal shape of the distribution, and (ii) a fat tail for longer links. The main differences between the curves obtained for different node degrees are a decreasing importance of link lengths in the ``plateau region'' above about 30 m, and a successive decrease of the maximum value of $l_l$ with increasing $k_n$. Both findings contribute to the observed decrease of the average link length with increasing node degree (Tab.~\ref{tab:German_nkP}). Our corresponding results confirm recent findings of Masucci \textit{et~al.} \cite{Masucci2009} obtained for the London street network. 

We note that the observed predominance of short links coincides with the typical size of building blocks in densely urbanised areas (50-100 m). The plateau in the frequency distribution is probably related to an inefficiency in having very short road sections especially in parts of the city with high traffic load, which would result in frequent stops at intersections. This hypothesis is supported by the position of the (local) maximum of $p(l_l)$ for $k_n=3,4$ (possibly highlighting the mainly rectangular network structures).

\subsection{Link angle distributions}

In the following, we will discuss the distribution of link angles, which encode additional information regarding the vectorial properties of road segments in geographical space. We note that the meaningful definition of link angles requires the presence of a planar network, which is assumed to be the case in urban road systems. 

Let ${\cal E}_n(i)$ denote the set of undirected links that start or end at a node $i$ of a planar network. In this case, every link $j\in {\cal E}_n(i)$ is uniquely defined by three properties: (i) the location of node $i$, (ii) the link length $l_{l,j}$, and (iii) the angle $\psi_{l,j}$ with respect to a fixed reference axis (or, alternatively, its unit vector). For convenience, this angle will be counted following the standard mathematical convention (i.e., counter-clockwise). For further statistical treatment, we will consider the following definition:\\

\noindent \textit{Definition 1:} Under the conditions specified above, $j_r\in{\cal E}_n(i)$ is called \textit{right neighbour} of $j$ iff 
\begin{equation}
j_r=\mbox{arg}\min\left\{ \psi_{l,j'}-\psi_{l,j}\mod 2\pi | j'\in\mathcal{E}_n(i) \right\}.
\end{equation}
\noindent
In a similar way, $j_l\in{\cal E}_n(i)$ is called \textit{left neighbour} of $j$ iff 
\begin{equation}
j_l=\mbox{arg}\min\left\{ \psi_{l,j}-\psi_{l,j'}\mod 2\pi | j'\in\mathcal{E}_n(i) \right\}.
\end{equation}
\noindent
The \textit{link angle} $\theta_{l,j}$ is then, for convenience, defined as the planar angle between the link $j$ and its right neighbour $j_r$, i.e.,
\begin{equation}
\theta_{l,j}=\psi_{l,j}-\psi_{l,j_r}\mod 2\pi.
\label{def:la}
\end{equation}

The link angle directly relates the spatial distribution of the nodes (in particular, their degrees) with the statistical features of links, which is a specific characteristic of planar networks. Unlike the link lengths, link angles exclusively refer to the respective node since 
\begin{equation}
\sum_{j\in{\cal E}_n(i)} \theta_{l,j}=2\pi.
\end{equation}

\begin{figure}[t!]
\begin{center}
\resizebox{0.48\columnwidth}{!}{\includegraphics{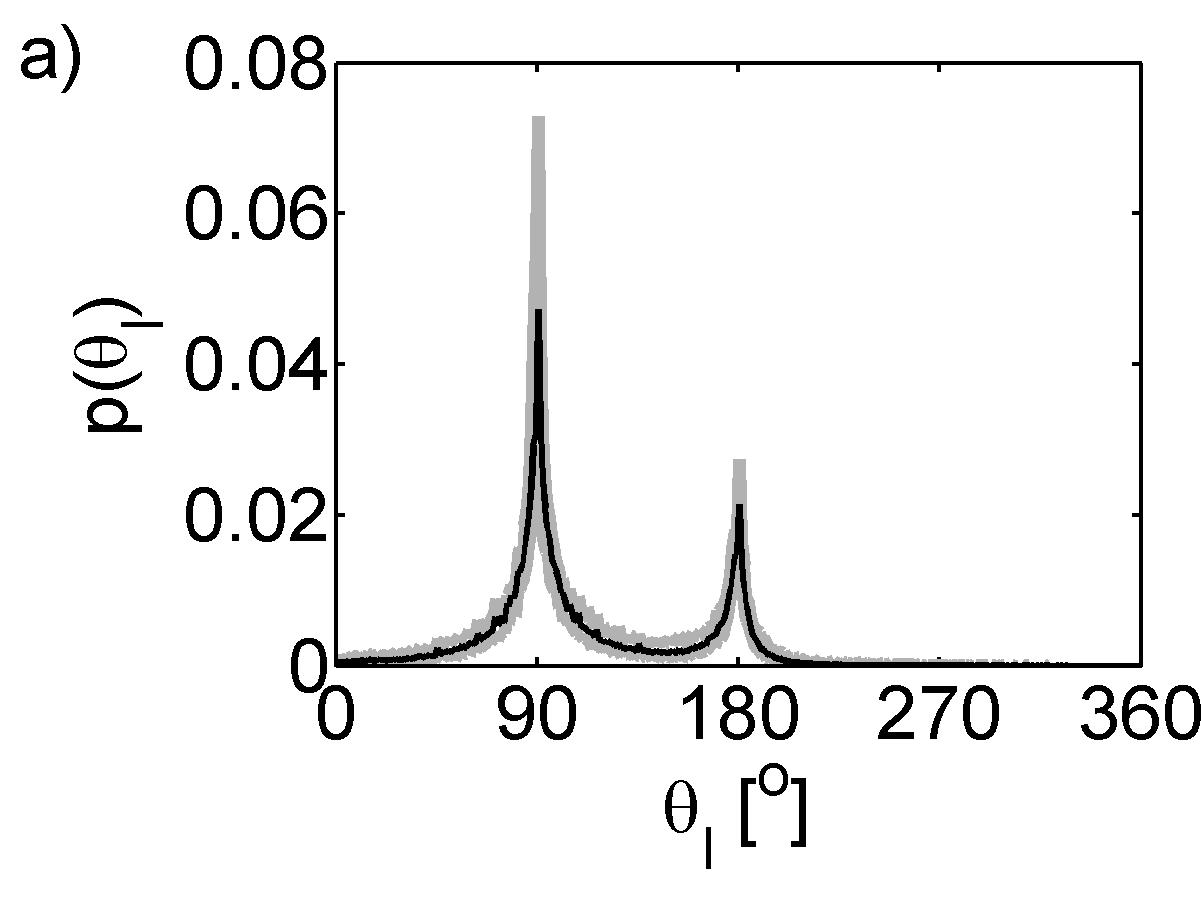}} \hfill
\resizebox{0.48\columnwidth}{!}{\includegraphics{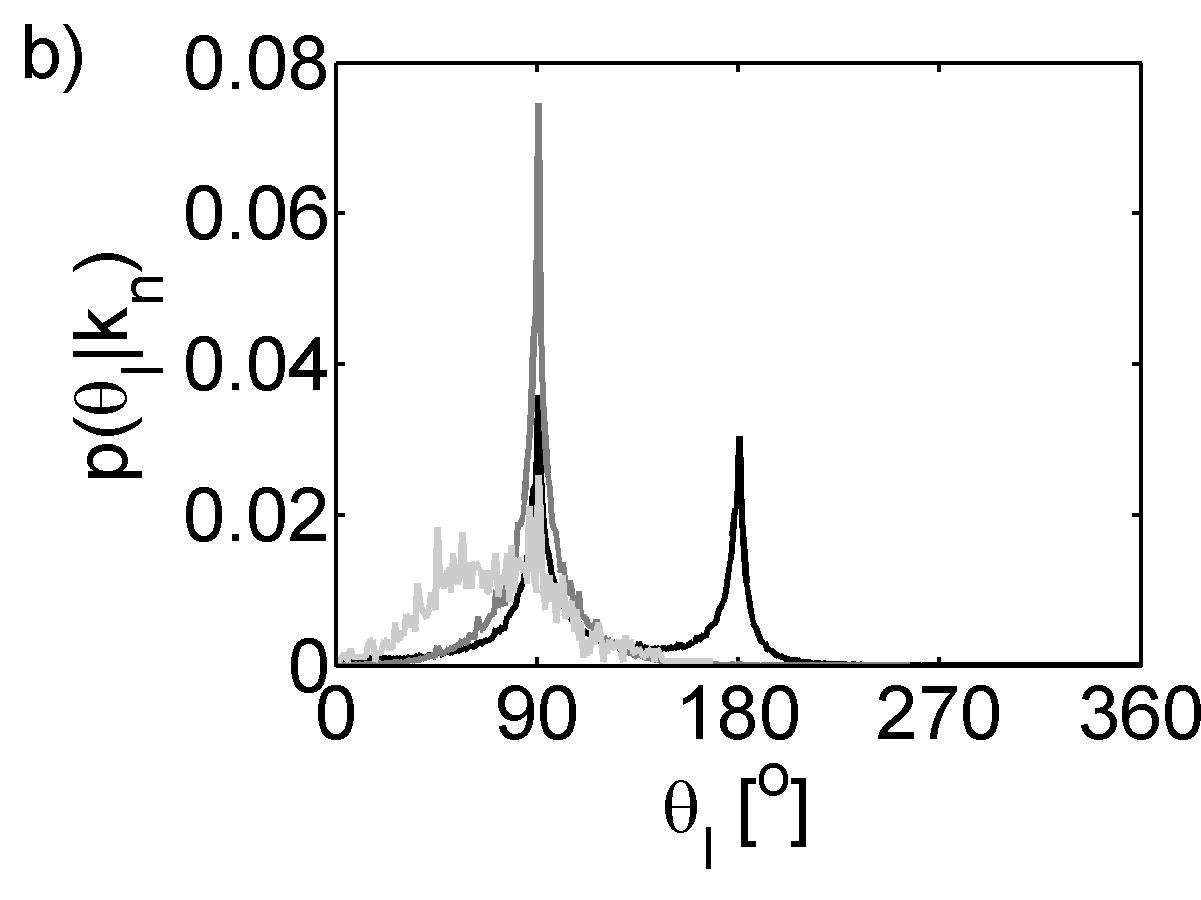}} \\
\resizebox{0.48\columnwidth}{!}{\includegraphics{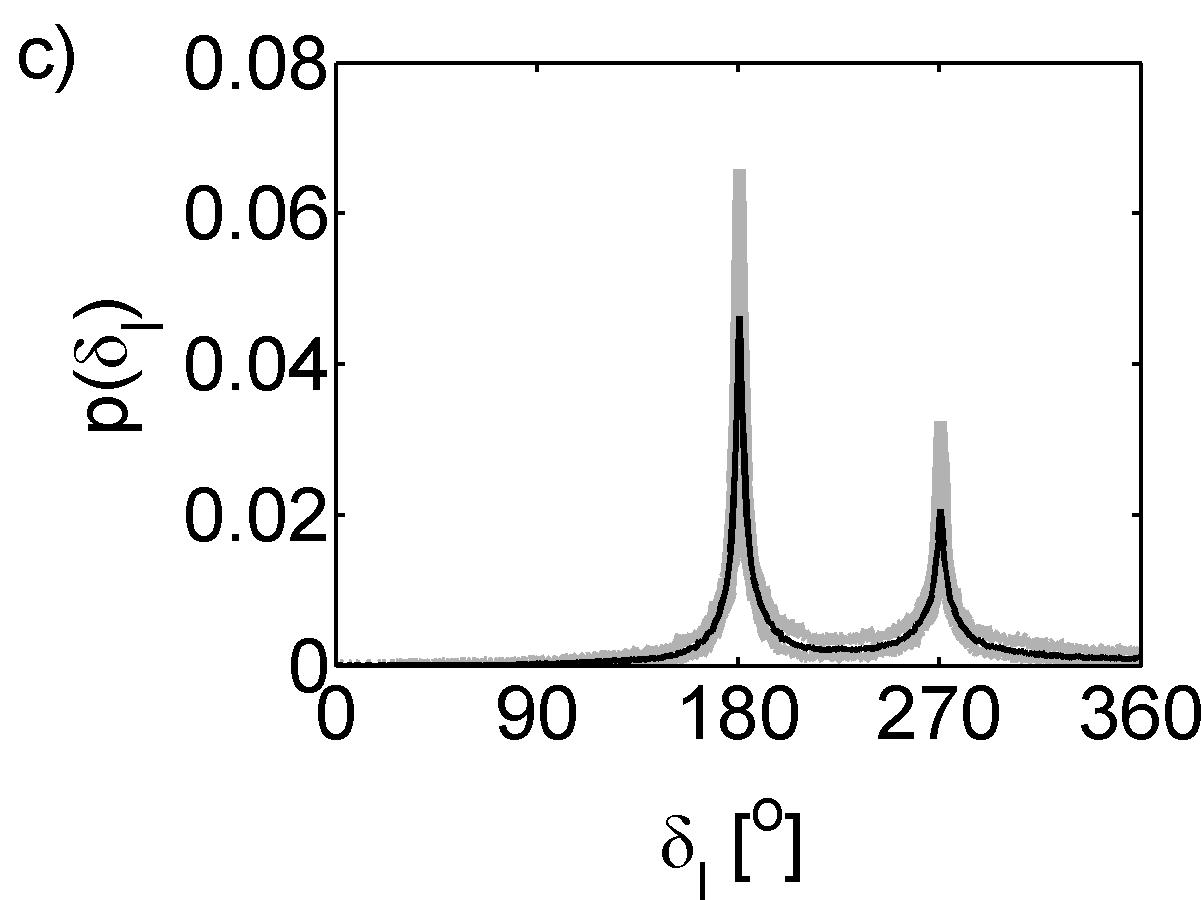}} \hfill
\resizebox{0.48\columnwidth}{!}{\includegraphics{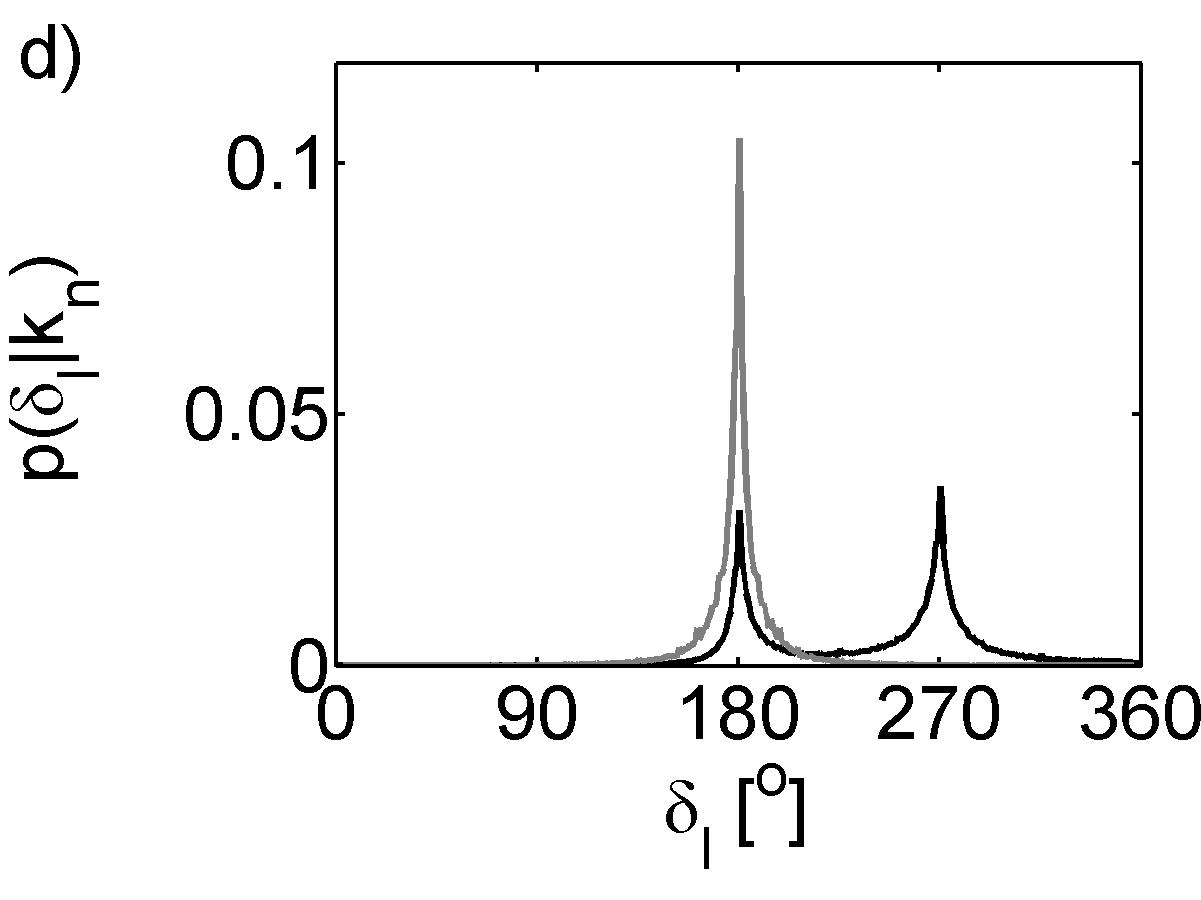}} \\
\resizebox{0.48\columnwidth}{!}{\includegraphics{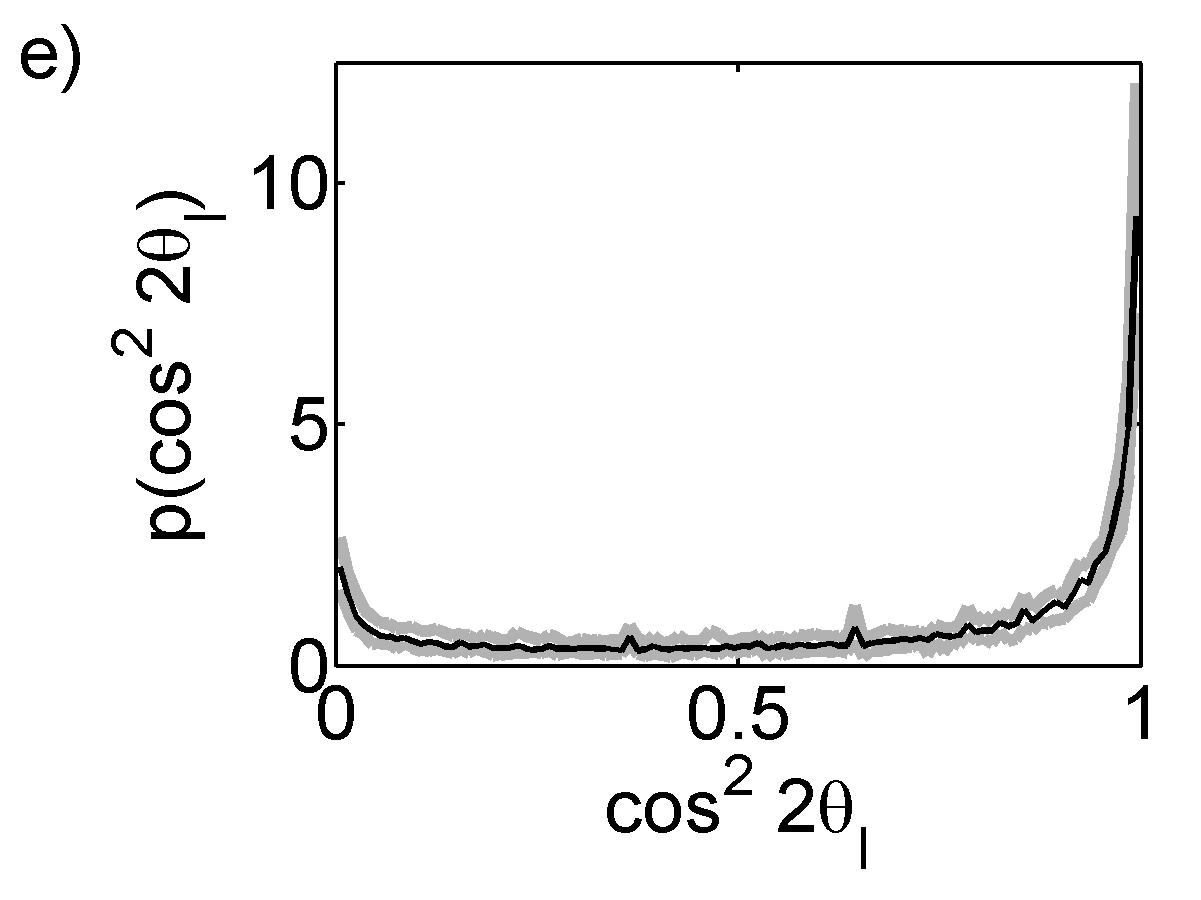}} \hfill
\resizebox{0.48\columnwidth}{!}{\includegraphics{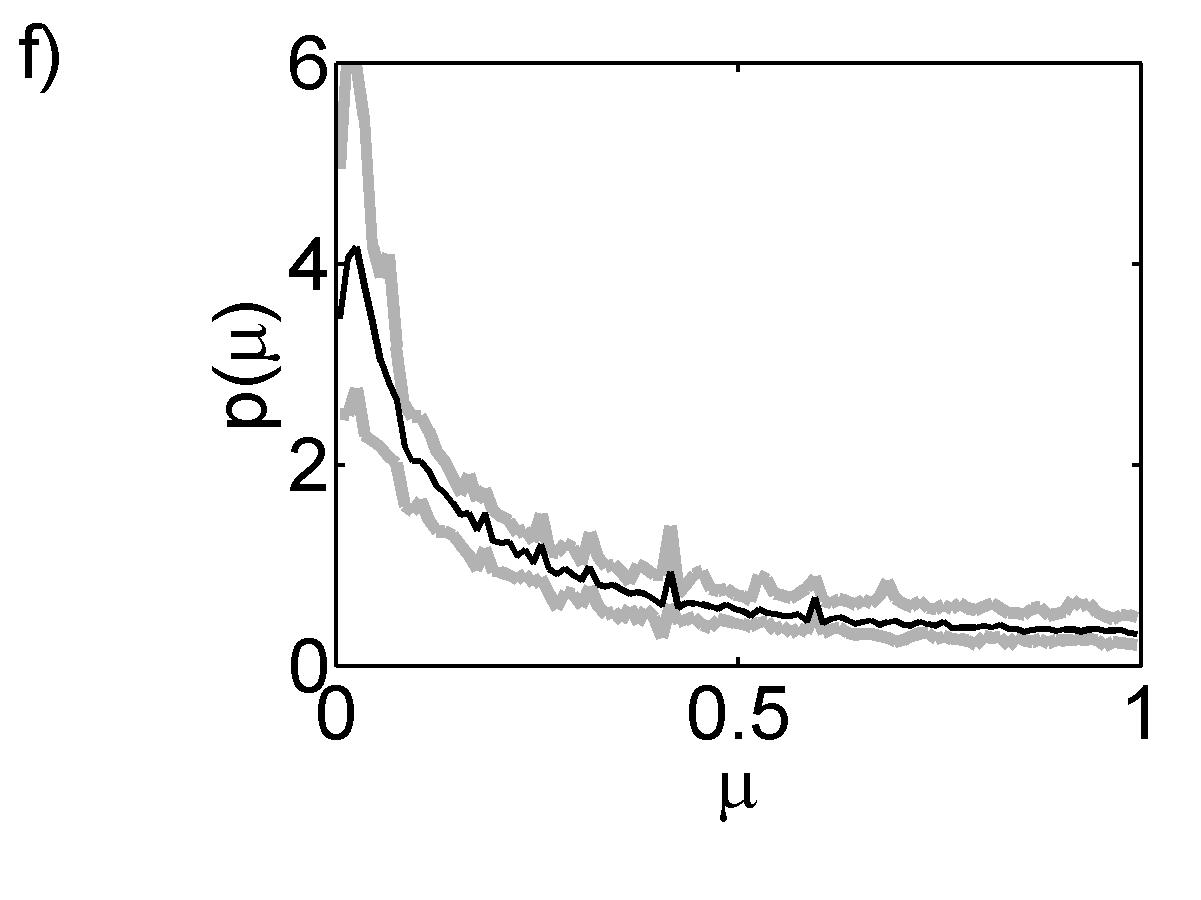}} \\
\end{center}
\caption{a) Average frequency distribution $p(\theta_l)$ of the link angles from the processed data taken over all cities (black), and upper and lower limits obtained from the individual cities (grey). b) Conditional distributions $p(\theta_l|k_n)$ for node degrees $k_n=3$, 4 and 5 (from black to light grey). c,d) Same as a,b) for the double angle distribution $p(\delta_l)$ (only $k_n=3,4$ are shown in d). e,f) Distributions of the parameters $\cos^2 2\theta_l$ (e) and $\mu$ (f) measuring the local deviation from a perfectly rectangular profile. The black line again represents the mean taken over all cities, grey lines indicate the minimum and maximum values for the individual road networks.}
\label{fig:la}
\end{figure}

As for the link lengths, all 20 road networks considered in this study show again approximately the same frequency distribution of the link angles as displayed in Fig.\ \ref{fig:la}a. This universal behaviour supports the hypothesis of a common construction principle of all studied urban road networks. Indeed, the mean link angles for the 20 individual cities scatter by far less than the link lengths (i.e., between 108.6$^o$ (Berlin) and 113.1$^o$ (Bremen), see Tab.~\ref{tab:GermanChar12b}). Note that there is an almost perfect (negative) statistical dependence between the mean link angle and the mean node degree $\left<k_n\right>$, since $\left<\theta_{l,j}\right>_j=360^o/k_n$.

In general, the link angle distributions $p(\theta_l)$ have two pronounced peaks at $90^o$ and $180^o$, where link angles of about $90^o$ are more abundant and show a larger dispersion than those of about $180^o$. This observation is directly related to the node degree distribution discussed above: When considering only nodes of degree $k_n=4$, there is only one sharp peak at $90^o$ (see Fig.\ \ref{fig:la}b), which corresponds to almost perpendicular intersections of pairs of roads as an extraordinarily probable way of forming nodes. For $k_n=3$, the link angles around $90^o$ occur twice as often as those around $180^o$ (note that the ratio between the corresponding peak frequencies in Fig.~\ref{fig:la}b is clearly smaller than 2, since the dispersion around the 90$^o$ peak is larger), which implies that the corresponding nodes are mainly formed by a straight primary road from which a secondary one splits perpendicularly.

In order to better understand the observed behaviour, remember that in the construction of urban road networks, minimisation of construction and maintenance costs as well as maximum usage of the road have to be taken into account \cite{Gastner2006}. In terms of geometry, the shortest Euclidean connection is the best solution for constructing the new road at minimum costs. This shortest line drawn from a node to an existing link naturally forms a perpendicular intersection. If this new road is further continued, four right angles are formed. If it touches only, two right angles and one $180^o$ angle emerge. Moreover, in terms of practical usage of the available space (in particular, regarding the typical building geometry), rectangular shapes are preferable. Finally, also in terms of turning vehicular traffic, symmetric 90$^o$ angles are a particularly reasonable choice, since for $\theta_l\not\approx 90^o$, turning processes become much more complicated in terms of curve radius and speed. The link angle distribution strongly support this explanation, which identifies a main driving mechanism for growing urban road networks. 

For a quantitative evaluation of the rectangularity of a given road network, we consider some statistics that measure the local deviation from a perfectly rectangular grid. As two properly normalised parameters, the squared cosine of the doubled link angles $\cos^2 2\theta_l$ and the normalised minimum absolute deviation from right angles
\begin{equation}
\mu=\frac{4}{\pi} \min_{k\in\mathbb{Z}} \left\{ \left| \theta_l-\frac{k}{2}\pi \right| \right\}
\end{equation}
\noindent
have been computed for all link angles. The mean values and associated standard deviations for all cities are listed in Tab.~\ref{tab:GermanChar12b}. According to these values, Munich, Mannheim and Berlin have the ``most rectangular'' road networks. In turn, the road networks of Wuppertal and Bochum show the least pronounced perpendicular splitting of roads. However, all values still lie within a relatively small range. As expected, both measures strongly correlate ($r=-0.981$, $\rho=-0.985$ for the respective mean values of the individual cities (in a perfectly rectangular network, we have $\left<\cos^2 2\theta_l\right>=1$ and $\mu=0$ by definition). However, if both characteristics are computed for all nodes in the studied networks, we observe a significant deviation from the ideal case, which results in a broadening of the corresponding probability distribution functions (Fig.~\ref{fig:la}e and f).

\subsection{Double-angle distributions}

In the link angle distributions, we have still observed deviations from the perfectly rectangular alignment of intersecting roads, which confirms daily life experience. Hence, our results obtained so far do not necessarily imply that intersections are also formed by \textit{straight} road segments. In order to further study the straightness feature explicitly, the joint distributions of neighbouring link angles are studied next. This analysis can be simplified by introducing the following auxiliary variable: \\

\noindent \textit{Definition 2:} Under the conditions of Definition 1, the \textit{double angle} $\delta_{l,j}$ associated with a link $j$ that enters a node of degree $k_n>2$ is defined as the angle between its left and right neighbours:
\begin{equation}
\delta_{l,j}=\psi_{l,j_l}-\psi_{l,j_r}\mod 2\pi.
\label{def:da}
\end{equation}

While the distribution of the individual link angles $\theta_l$ quantifies the \textit{rectangularity} of intersections, at least for nodes with degree $k_n=3$ and 4 (i.e., about 99.7\% of all intersections, see Tab.~\ref{tab:German_nkP}), the associated double angles $\delta_l$ measure the \textit{straightness} of the crossing roads. It is trivial to demonstrate that 
\begin{equation}
\delta_{l,j}=\theta_{l,j_l}+\theta_{l,j}. 
\end{equation}
\noindent

The distributions of double-angles (Fig. \ref{fig:la}c and d) show the behaviour to be expected from the link angles. In particular, for nodes with degree $k_n=4$, there is only one sharp peak around $180^o$, while for nodes with $k_n=3$, there are two peaks at $180^o$ and $270^o$. Note, however, that although one would expect double angles of about $270^o$ to occur twice as often as those of about $180^o$ for $k_n=3$, we observe that the corresponding peak frequencies differ less from each other. This again implies that the dispersion is larger around the $270^o$ maximum (see Fig.~\ref{fig:la}d). Note that contributions due to nodes with degrees $k_n\geq 5$ do not have any remarkable impact on the full double-angle distribution due to their very low abundance.

\subsection{Statistical model for link angle and double-angle distributions}

In the following, our above considerations will be supplemented by a simple statistical model for the observed link angle and double-angle distributions. In particular, for $k_n=3$ and 4, the sharp and partially symmetric peaked structure of the observed frequency distributions suggests that a stretched exponential (or generalised Laplacian)
\begin{equation}
p(\theta_l|\alpha,\tau)\simeq\frac{1}{N}\exp\left\{-\left|\frac{\theta_l-\theta_{l,0}}{\tau}\right|^{\alpha}\right\}
\end{equation}
\noindent
(or a similar expression for $\delta_l$) could provide a reasonable statistical model. In the latter expression, 
\begin{equation}
N=\frac{2\tau}{\alpha}\Gamma(1/\alpha) \quad \mbox{with} \quad \Gamma(t)=\int_{0}^{\infty} e^{-u} u^{t-1} du
\end{equation}
\noindent
is a normalisation constant, $\theta_{l,0}$ the peak location, $\tau$ measures the dispersion of the distribution around the peak, and the exponent $\alpha$ mainly determines the scale of the probability decay as $\theta_l$ deviates from the peak position.

Stretched exponential distributions have been recently used for describing the \textit{cumulative} frequency distributions of a variety of natural as well as economic time series \cite{Laherrere1998} (e.g., peak-trough patterns of commodity prices \cite{Roehner1998}). There are only few examples for applications as models for probability \textit{densities}, e.g., of relative price increments (asset returns) in financial time series \cite{McCauley2003}. In a wider sense, generalised Laplacians have been used for approximating the distributions of wavelet coefficients in image processing applications \cite{Mallat1989}.

The full distributions of both link and double angles show minor asymmetries around their respective peaks, which is related to the presence of nodes of degree $k_n>4$. However, for $k_n=3,4$, one could expect a symmetric behaviour. Note that the full distributions $p(\theta_l)$ and $p(\delta_l)$ can be expressed as finite mixture models~\cite{Everitt1981} with components corresponding to different $k_n$, the statistical weights $\pi^{(k_n)}$ ($k_n=3,4,\dots$) of which are determined by the relative frequencies of the different node degrees $p(k_n)$, i.e., $\pi^{(k_n)}=p(k_n)\cdot k_n/\left<k_n\right>$.

\begin{figure}[t!]
\begin{center}
\resizebox{0.48\columnwidth}{!}{\includegraphics{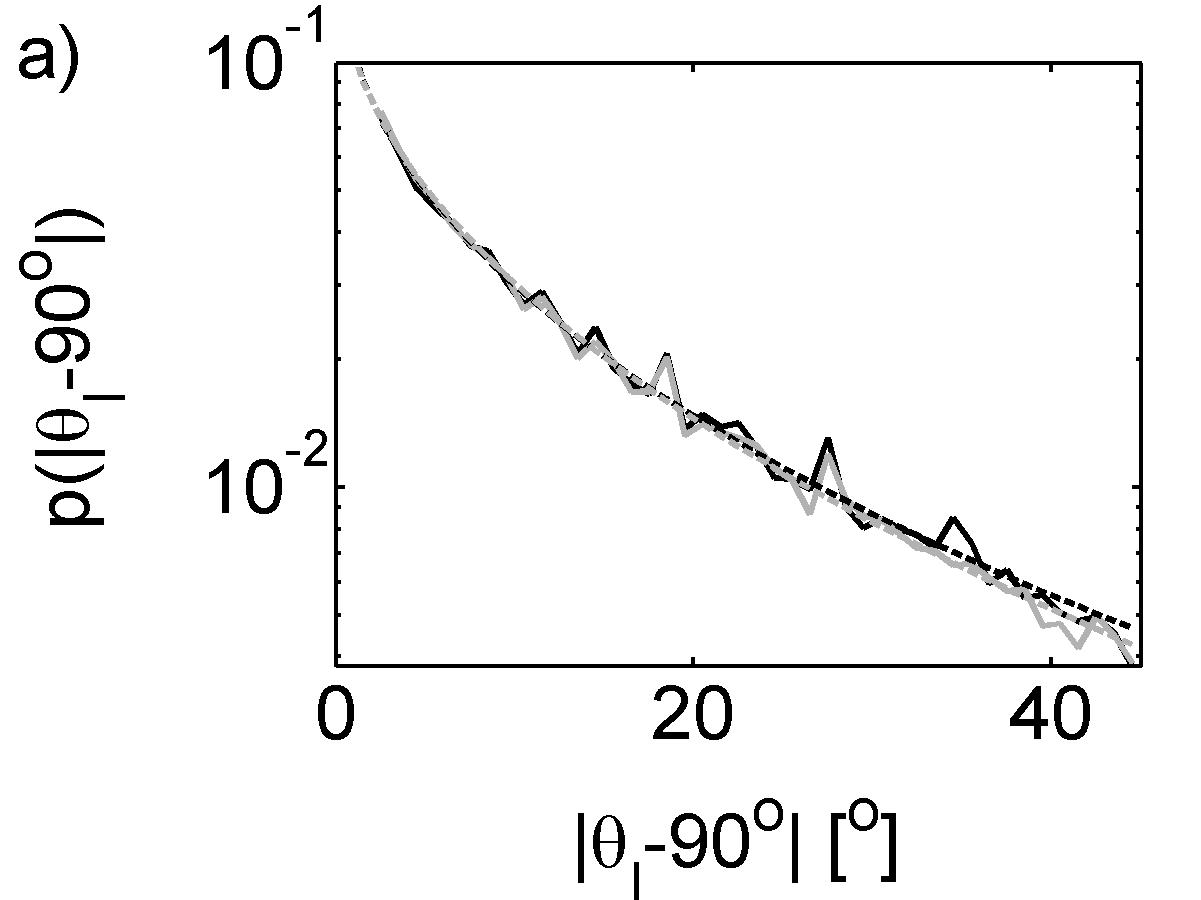}} \hfill
\resizebox{0.48\columnwidth}{!}{\includegraphics{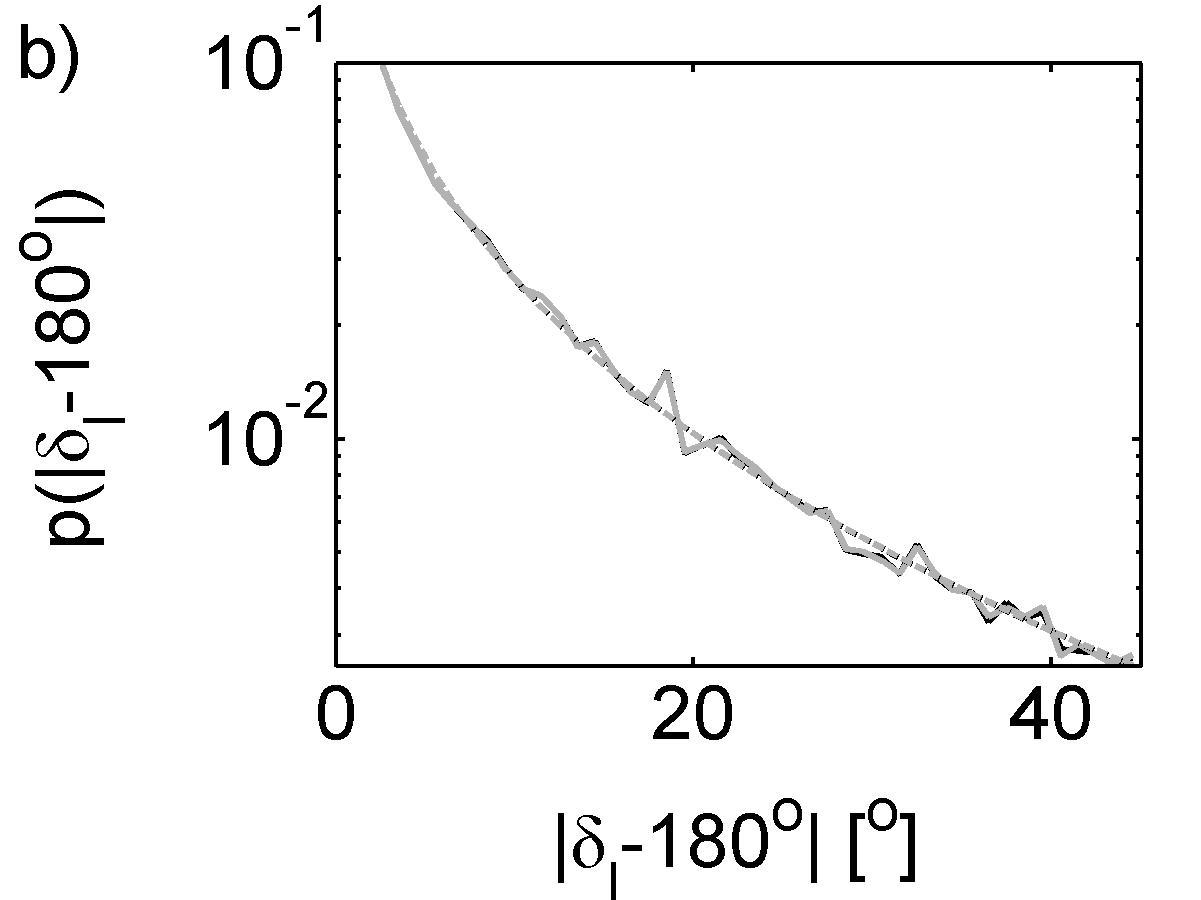}} \\
\resizebox{0.48\columnwidth}{!}{\includegraphics{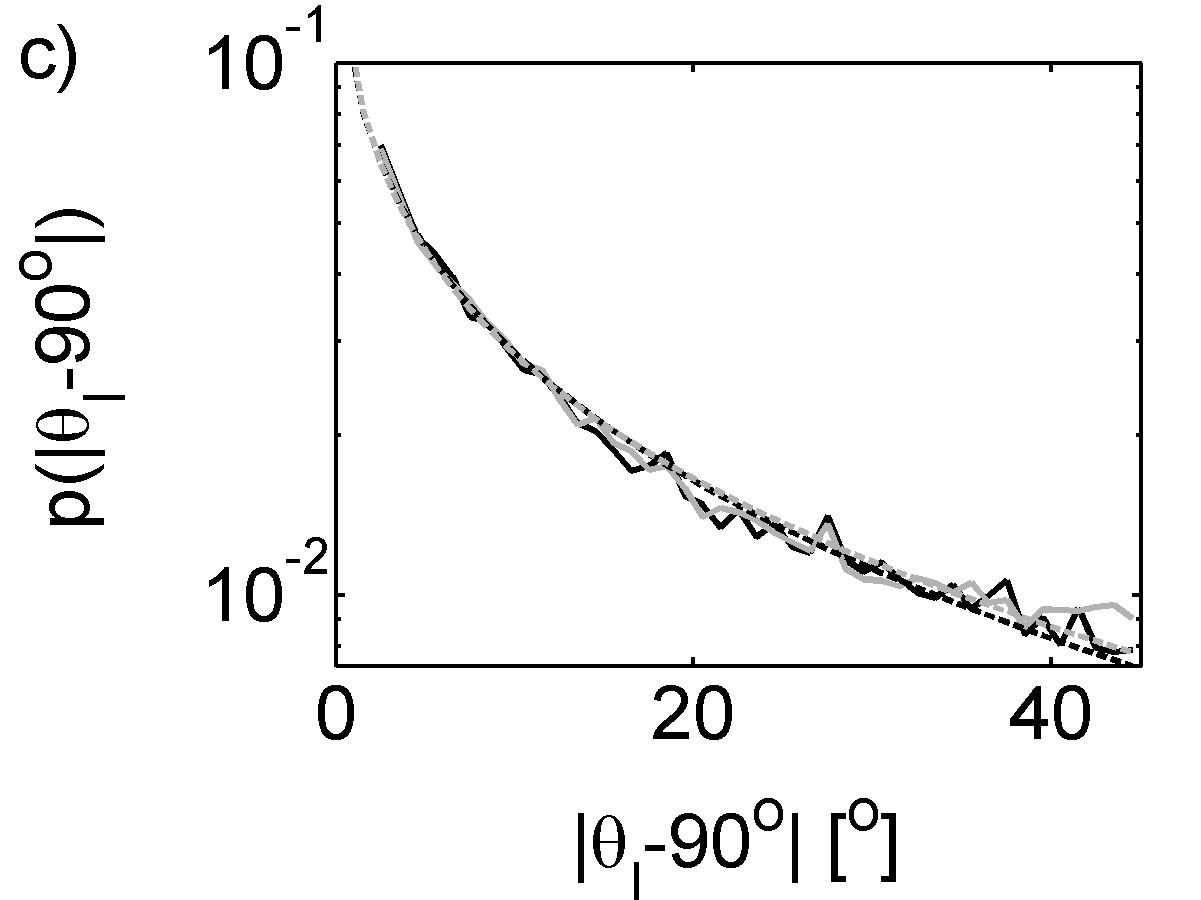}} \hfill
\resizebox{0.48\columnwidth}{!}{\includegraphics{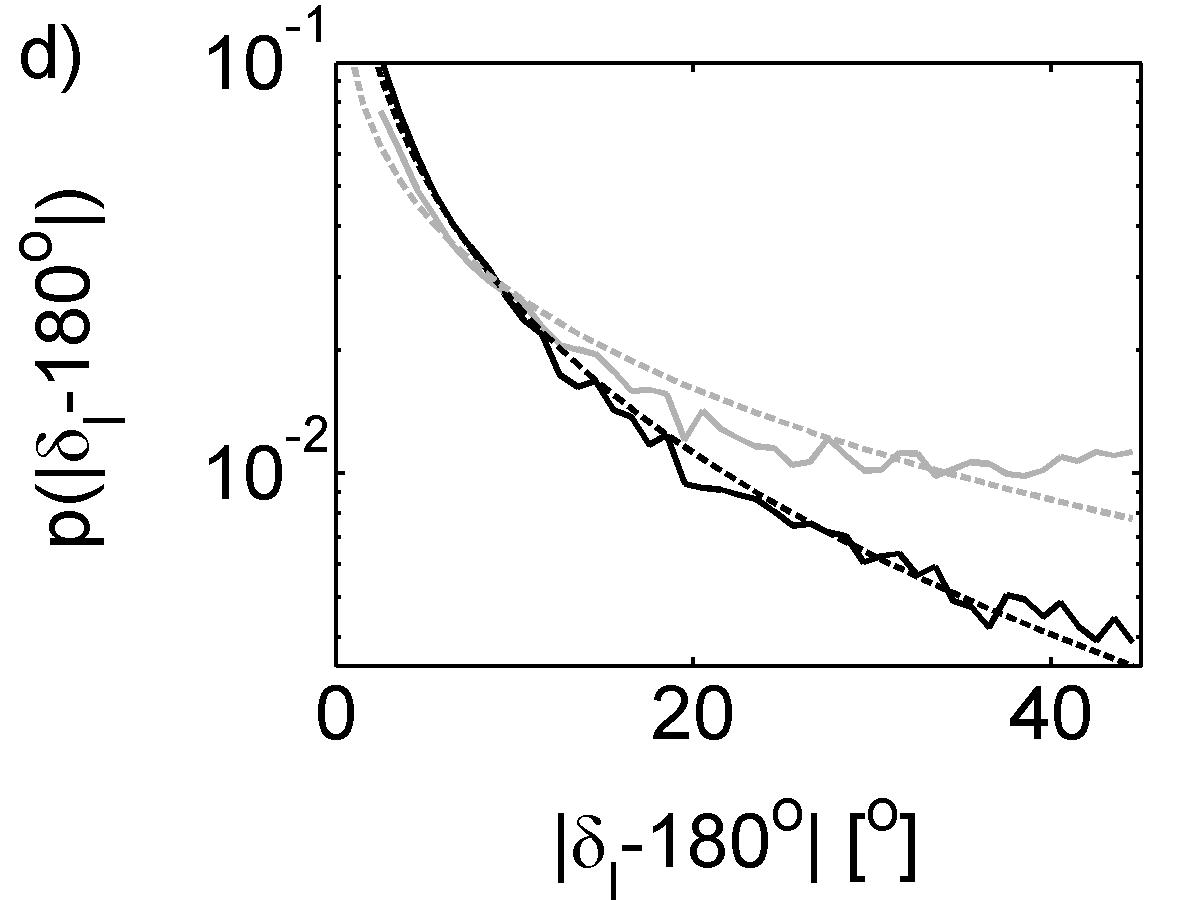}} \\
\resizebox{0.48\columnwidth}{!}{\includegraphics{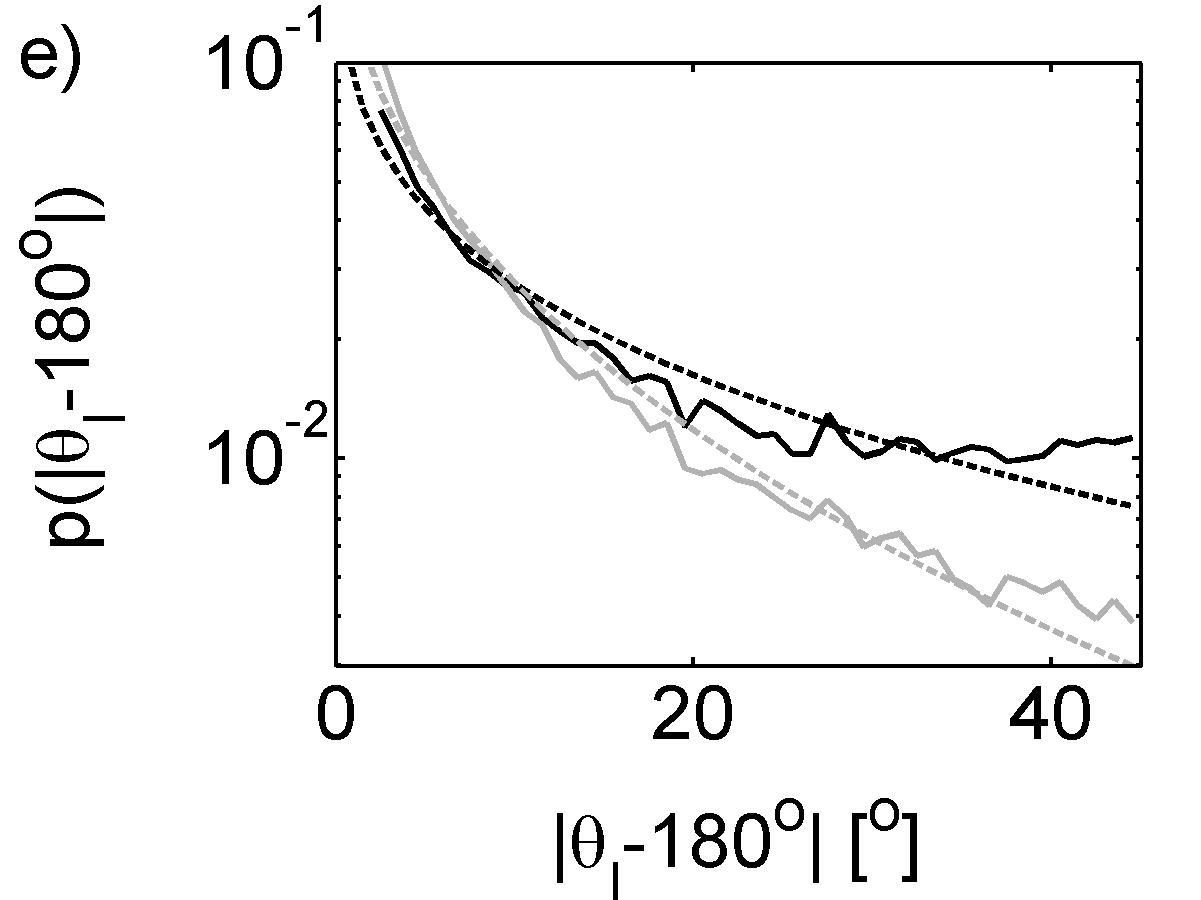}} \hfill
\resizebox{0.48\columnwidth}{!}{\includegraphics{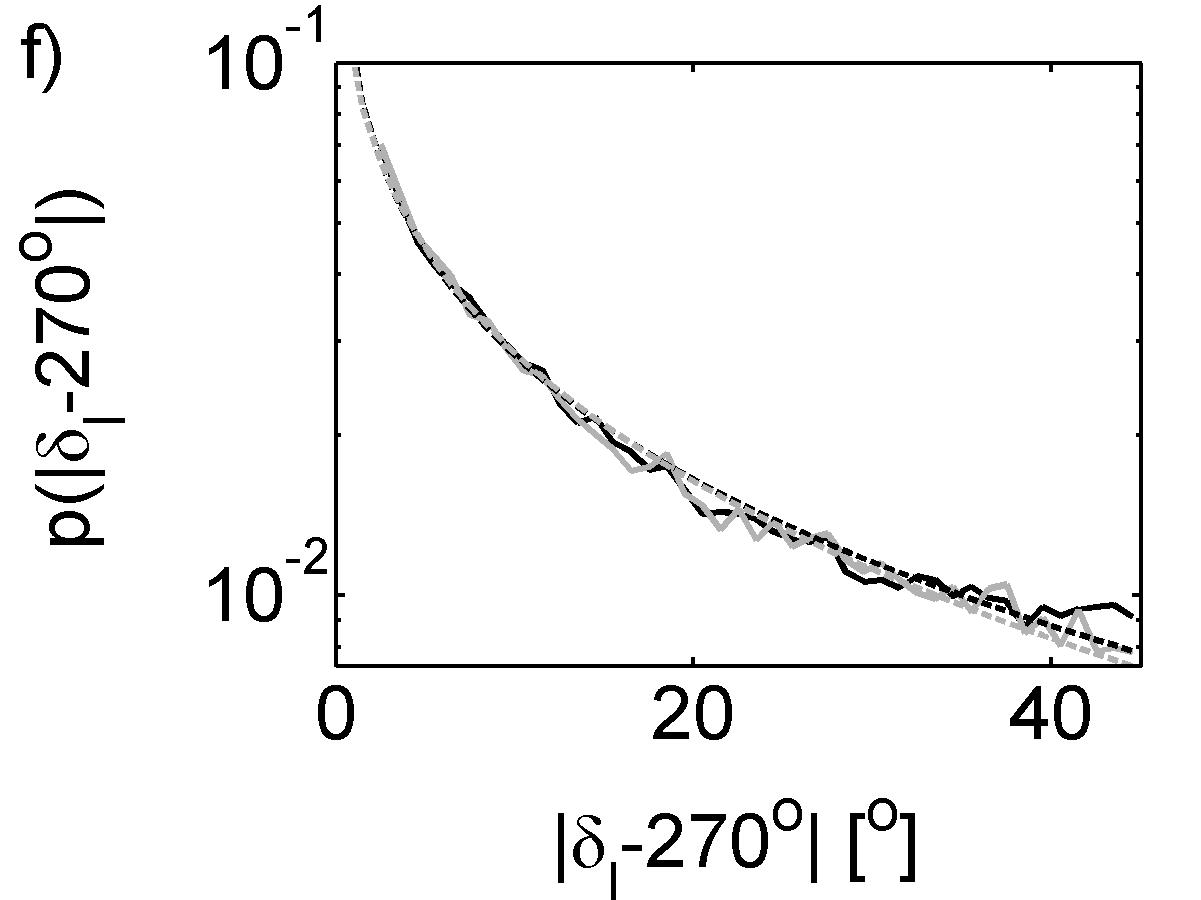}} \\
\end{center}
\caption{a) Empirical frequency distributions $p\left(\left|\theta_l-\theta_{l,0}\right|\right)$ (solid lines) for $k_n=4$ ($\theta_{l,0}=90^o$) at $\theta_l<\theta_{l,0}$ (black) and $\theta_l>\theta_{l,0}$ (grey), and corresponding stretched exponential fit (dashed lines). b) Same as a) for the double-angles $\delta_l$ ($\delta_{l,0}=180^o$). c,d) Same as a,b) for $k_n=3$. e,f) Same as c,d) for $\theta_{l,0}=180^o$ and $\delta_{l,0}=270^o$, respectively. Note that some of the lines are hardly visible, since they coincide almost completely with others.}
\label{fig:lafit}
\end{figure}

For $k_n=4$, the double angle distributions for $\delta_l<180^o$ and $\delta_l>180^o$ pass different statistical tests for homogeneity (Kolmogorov-Smirnov, Kuiper, $\chi^2$) on a 99\% confidence level. For the corresponding link angle distributions, tests on the explicit data (Kol\-mo\-go\-rov-Smirnov, Kuiper), however, reject the null hypothesis of symmetric behaviour around $\theta_l=90^o$. In contrast, since rounding errors in the data base could matter here, binning of the corresponding data seems reasonable, which leads (for a binwidth of $0.5^o$) to $\chi^2$ values that do not allow rejecting the null hypothesis of symmetric tails anymore. Figure~\ref{fig:lafit} demonstrates this symmetry and underlines that at least for $\left|\theta_l-90^o\right|\gtrsim 2^o$ and $\left|\delta_l-180^o\right|\gtrsim 2^o$, stretched exponentials fit the behaviour of the empirical distribution functions reasonably well. Note that the explicit estimation of the corresponding model parameters is a challenging task, which is beyond the scope of the presented study. However, as a general result, we summarise that both the associated scaling exponents $\alpha$ and the dispersion parameters $\tau$ are clearly larger for the link angles than for the double angles. This implies a broader distribution near the peak, but a faster decay in the tails of $p(\theta_l)$ (see Fig.~\ref{fig:la}).

In contrast to the previously considered case, for $k_n=3$ the link and double-angle distributions have two peaks with different properties. Although the distributions of the tails associated with these peaks do not pass the homogeneity tests mentioned above, it can be seen from Fig.~\ref{fig:lafit} that there is a considerable degree of symmetry around the $90^o$ peak of $p(\theta_l)$ and the $270^o$ peak of $p(\delta_l)$, respectively, whereas the tails associated with the $180^o$ peaks show strong asymmetries. Similar statements hold for the possibility of describing the empirical distributions by stretched exponential functions. The observed asymmetry around the 180$^o$ peak is possibly related to the mutual overlap of the tails associated with different peaks, which cannot be sufficiently resolved by the heuristic cutoffs considered here in the middle between two peaks. A more sophisticated statistical treatment might help improving the corresponding results in future studies.

\subsection{Radial distributions of link properties}

The decrease of the typical link lengths with increasing degree of the involved nodes suggests a possible relationship with the spatial distribution of nodes and links within the studied cities. Even though a detailed investigation of the historical development of the 20 networks is beyond the scope of this study, in the following, we will approximate the temporal evolution by the distance of an intersection from the geometric city centre $\bar{\vec{x}}$. Although this approximation is only reasonable for symmetrically grown cities, we argue that historical quarters are typically located close to the geometric city centres, whereas modern parts can more often be found in the outer regions.

\begin{figure}[t!]
\begin{center}
\resizebox{0.48\columnwidth}{!}{\includegraphics{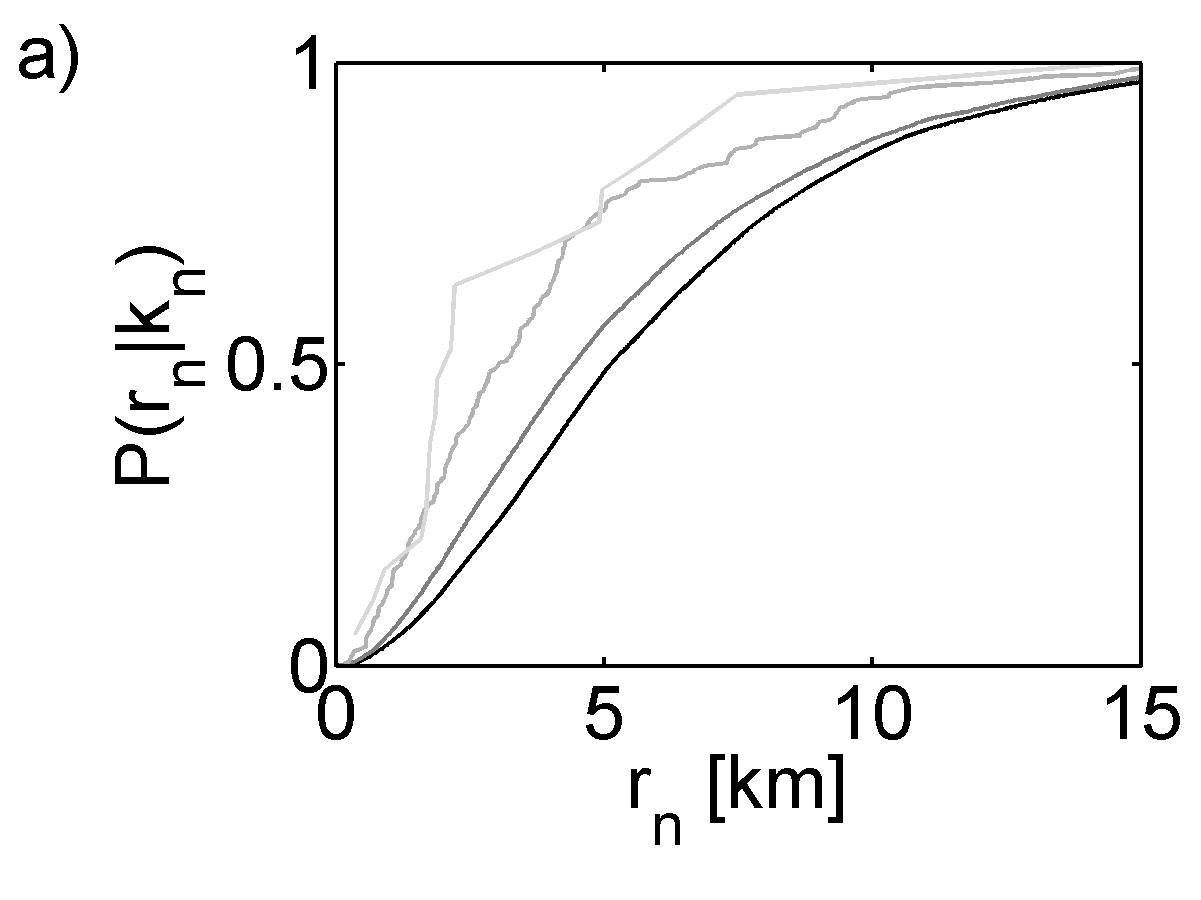}} \hfill
\resizebox{0.48\columnwidth}{!}{\includegraphics{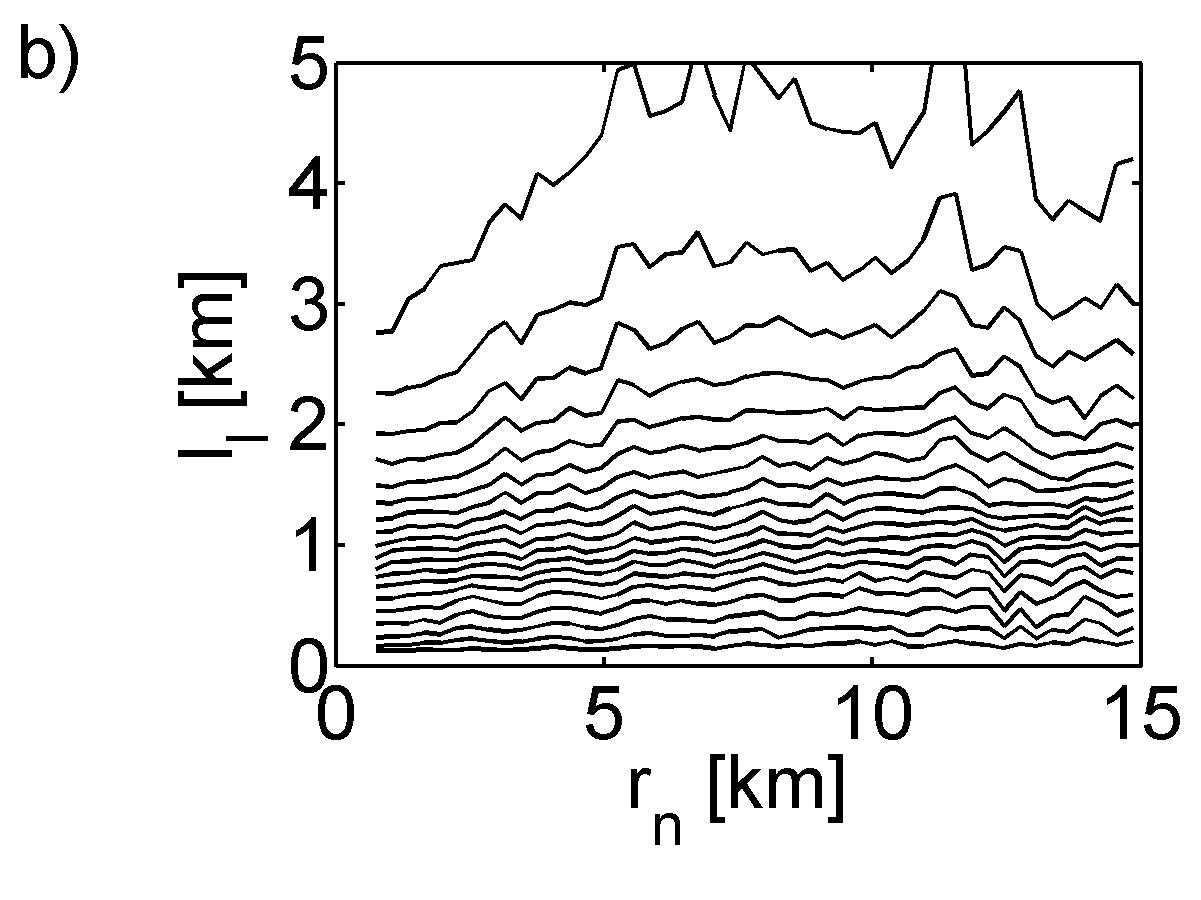}} \\
\resizebox{0.48\columnwidth}{!}{\includegraphics{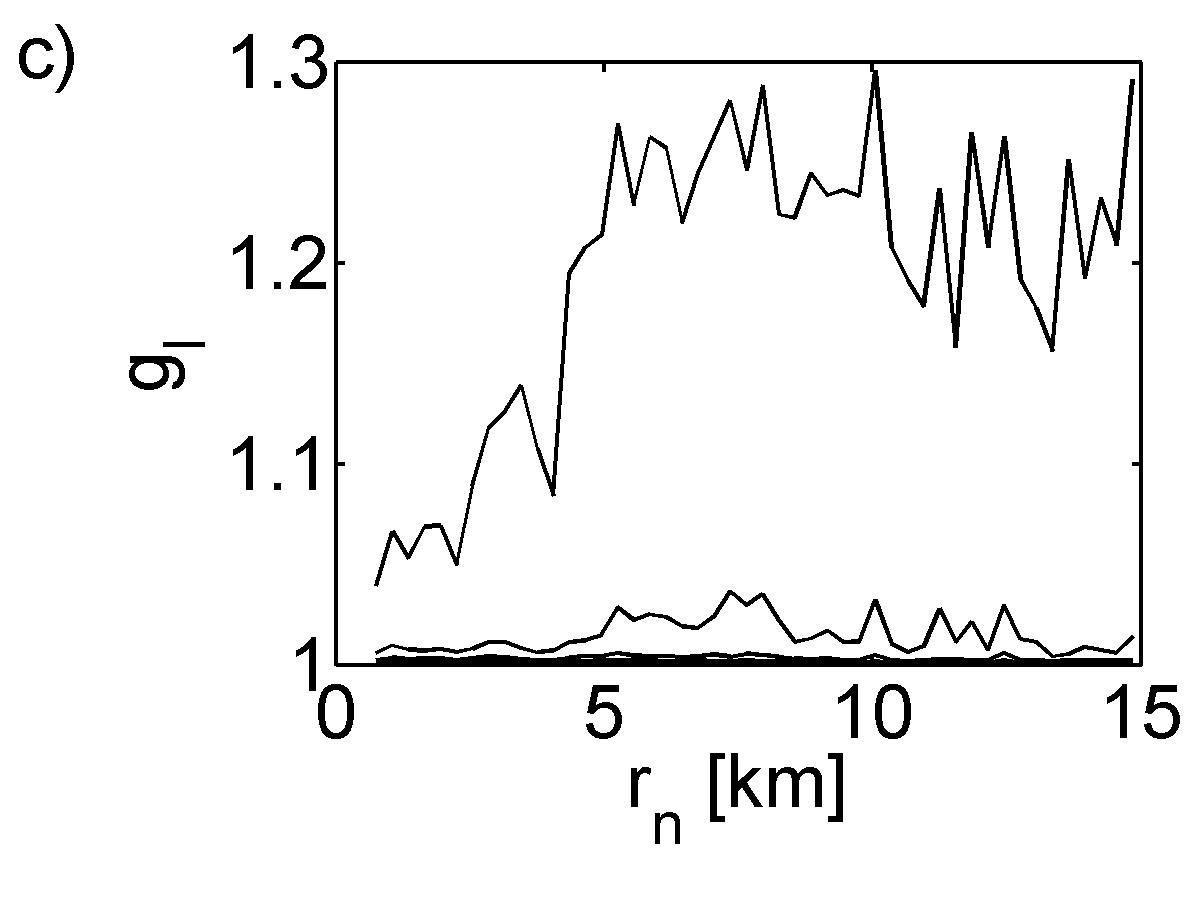}} \hfill
\resizebox{0.48\columnwidth}{!}{\includegraphics{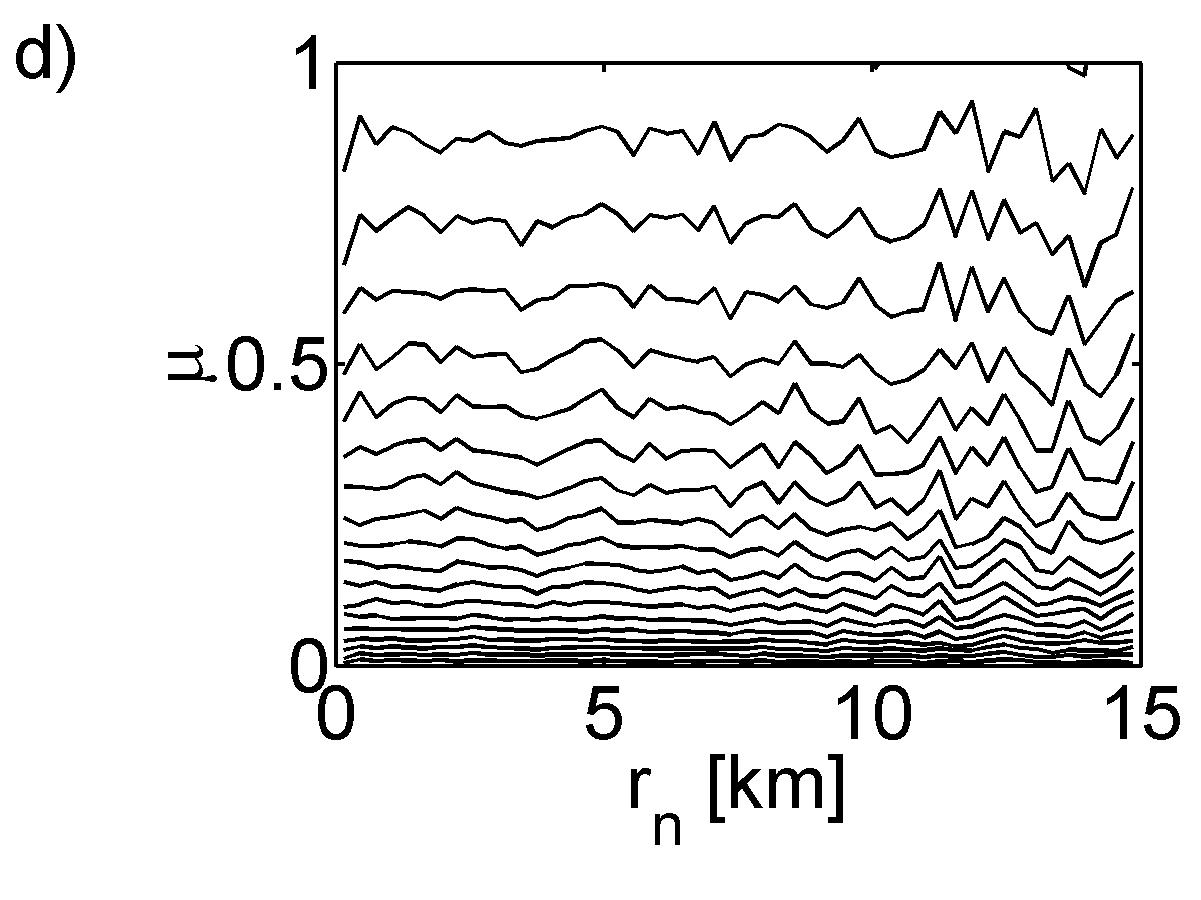}} \\
\end{center}
\caption{a) Conditional cumulative frequency distributions $P(r_n|k_n)$ of node distances from the estimated city centres for $k_n=3$, 4, 5, and 6 (from black to light grey). b)-d) 95\% to 5\% (in steps of 5\%, from top to bottom) probability levels of the empirical conditional distribution functions of the link lengths (b), the curvature parameter $g_l$ (c), and the rectangularity parameter $\mu$ (d) in dependence on the distance $r_n$ from the city centres.}
\label{fig:ll2}
\end{figure}

We speculate that nodes with high degrees are usually located in densely urbanised areas, in particular, more often relatively close to the city centres. In a similar spirit, we also expect shorter streets being typical for historical city centres. In order to examine this hypothesis, for every city, the distances of all nodes from the respective geometric centre have been computed. Fig.~\ref{fig:ll2}a demonstrates that there is indeed a tendency that high-degree nodes can be found more often (in terms of relative frequencies) close to the city centres than such with lower degrees. This result is again in good agreement with recent findings by Masucci \textit{et~al.} for the London street network, who observed an approximately linear decrease of the mean node degree with an increasing distance from the city centre \cite{Masucci2009}. In a similar way, the quantiles of the probability distributions of the link lengths show a clear tendency towards shorter links near the city centres (Fig.~\ref{fig:ll2}b), which also supports the results for the London street network \cite{Masucci2009}. However, while longer road segments occur especially often at intermediate distances (about 5-10 km) from the city centres, in the outer parts of the examined cities we again find a tendency towards a lower frequency of very long links, which could be related to a predominance of residential areas in these regions, which are typically characterised by short roads.

In addition, we evaluate a possible dependence of the road geometry on the distance from the city centres. As a parameter describing the curvature of individual road segments, we consider the ratio between the corresponding link length $l_l$ and the Euclidean distance between the two connected nodes,
\begin{equation}
g_l=\frac{l_l}{\| \vec{x}_{l,1}-\vec{x}_{l,2}\|}\geq 1.
\end{equation}
\noindent
Figure~\ref{fig:ll2}c shows that there is actually a tendency towards only few links with significant curvature ($g_l>1$) close to the city centres, which implies that in the corresponding parts of the cities, both short and straight road segments are particularly common. In contrast, there seems to be no clear trend towards more rectangular structures in these areas as shown by the corresponding distribution of the parameter $\mu$ in dependence on $r_n$ (Fig.~\ref{fig:ll2}d).

\section{Cell Properties} \label{sec:cells}

Roads divide urban areas into cellular substructures. The two-dimensional objects, which are formed by closed loops consisting of different links, are thus called the \textit{cells} of the road network \cite{Laemmer2006a}. Formally, let ${\cal V}_c(i)$ be the set of nodes that belong to a given cell $i$. The set ${\cal E}_c(i)$ of links forming the boundary of this cell is then given as
\begin{equation}
{\cal E}_c(i)=\bigcup_{m\in {\cal V}_c(i)}\left( \bigcup_{m'\in {\cal V}_c(i),m'\neq m} \left( {\cal E}_n(m) \cap {\cal E}_n(m') \right) \right).
\end{equation}

There is a variety of characteristic geometric properties describing a cell, in particular

\begin{itemize}
\item the cell area $A_c$,
\item the \textit{topological} cell degree $k_c$ (i.e., the number of neighbouring cells \cite{Godreche1992}, where neighbours are defined according to the presence of a common edge),
\item the \textit{geometric} cell degree $\kappa_c$ (i.e., the number of straight road segments forming the cell, note that Masucci \textit{et~al.} \cite{Masucci2009} called this quantity the cycle length $Cl$),
\item the (maximum) cell diameter $d_c$, and
\item the cell perimeter \cite{Barthelemy2008a,Barthelemy2008b}
\begin{equation}
p_c=\sum_{i\in {\cal E}_c} l_{l,i}.
\end{equation}
\end{itemize}
\noindent
It is obvious that the different measures are not mutually independent. For example, in order to characterise the shape of a cell, it is therefore convenient to combine cell area and diameter into one normalised parameter, the form factor (or structure factor) $\phi_c$ \cite{Laemmer2006a,Barthelemy2008a,Barthelemy2008b}, which is defined as the ratio between the actual cell area $A_c$ and the area of the smallest possible circumscribed circle. Since the latter one is determined by the maximum cell diameter as 
\begin{equation}
A_{min}(d_c)=\frac{\pi d_c^2}{4},
\end{equation}
\noindent
the defining equation of the form factor reads
\begin{equation}
\phi_c(A_c,d_c)=\frac{A_c}{A_{min}(d_c)}=\frac{4A_c}{\pi d_c^2}.
\end{equation}

In the following, the distributions of cell degrees, cell areas, and form factors will be studied as examples of the typical properties of urban road networks. Since we are interested in the physical shapes of the cells, the original data will be used here again instead of the processed ones. 

\begin{figure}[t!]
\begin{center}
\resizebox{0.48\columnwidth}{!}{\includegraphics{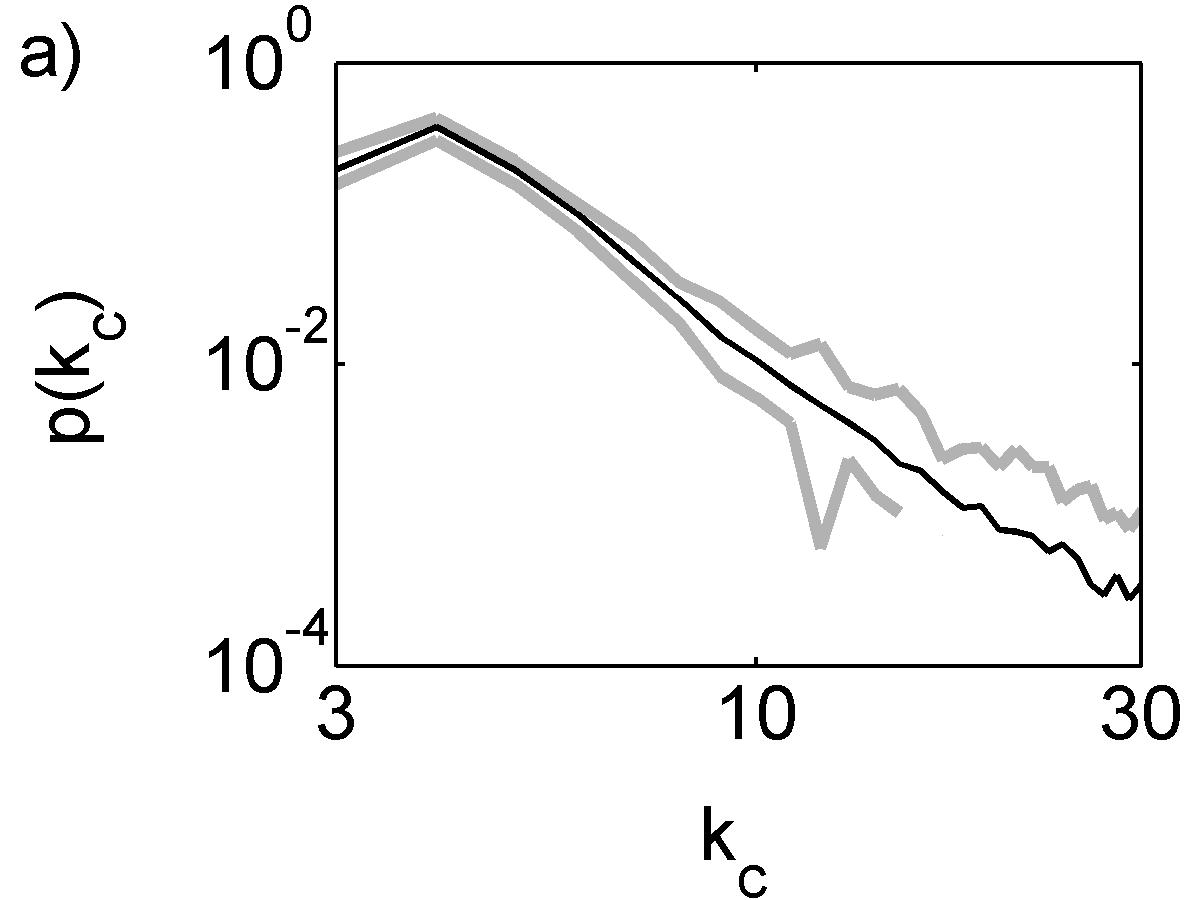}} \hfill
\resizebox{0.48\columnwidth}{!}{\includegraphics{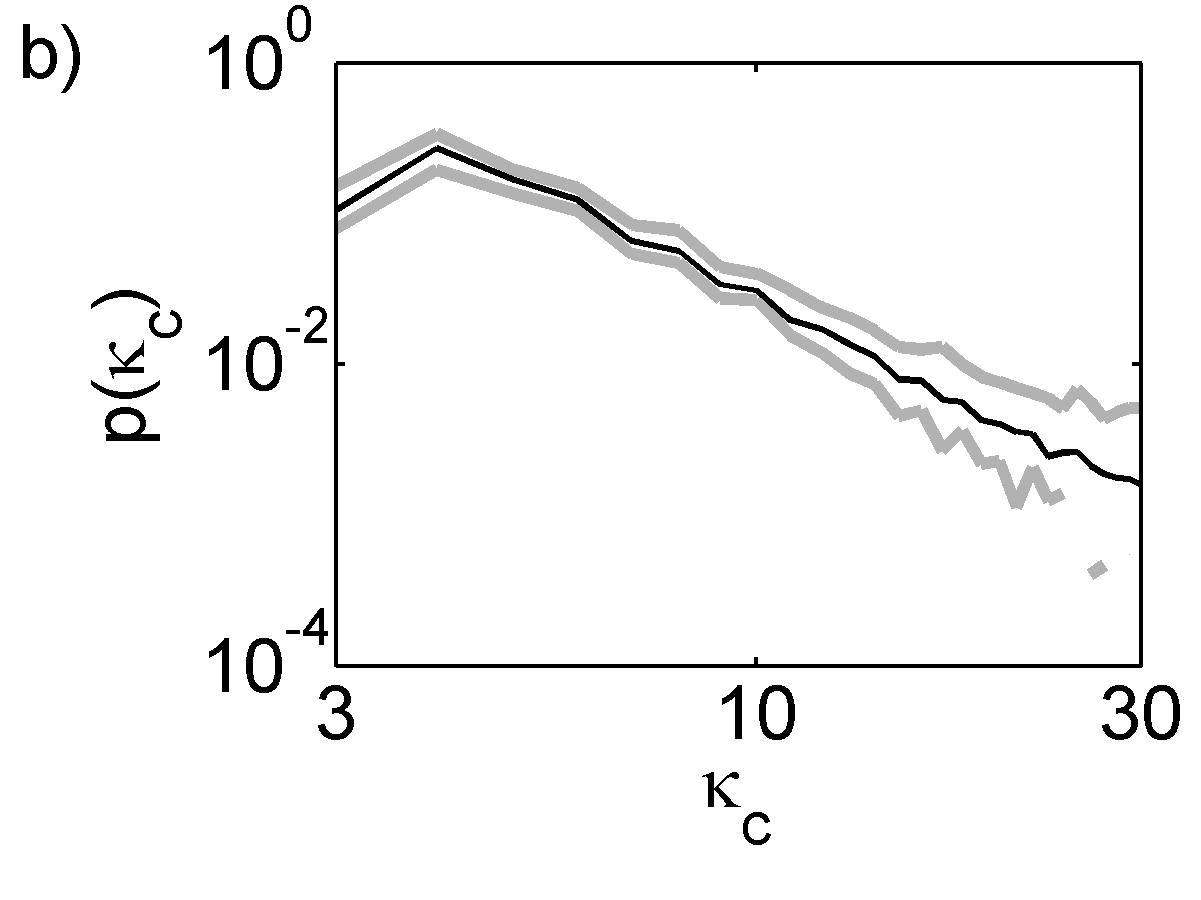}} \\
\resizebox{0.48\columnwidth}{!}{\includegraphics{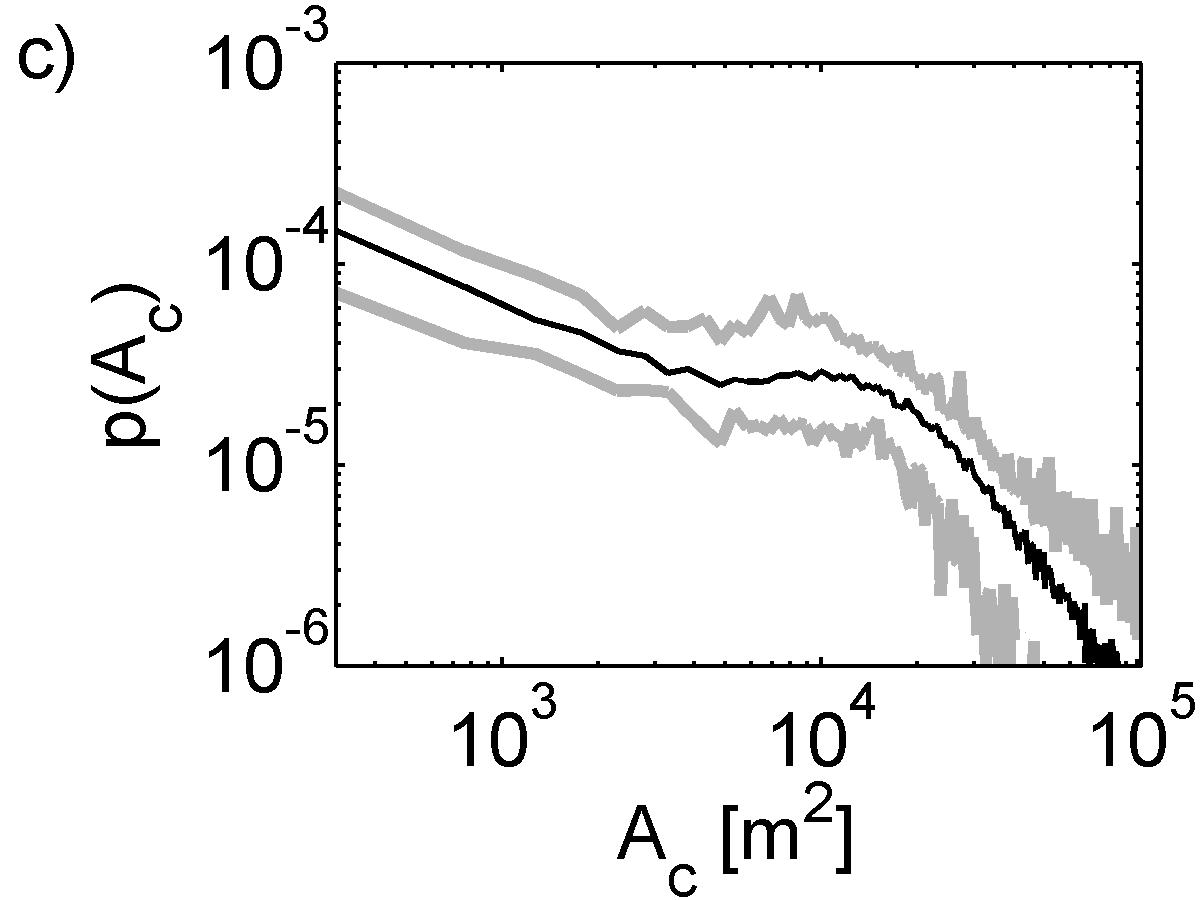}} \hfill
\resizebox{0.48\columnwidth}{!}{\includegraphics{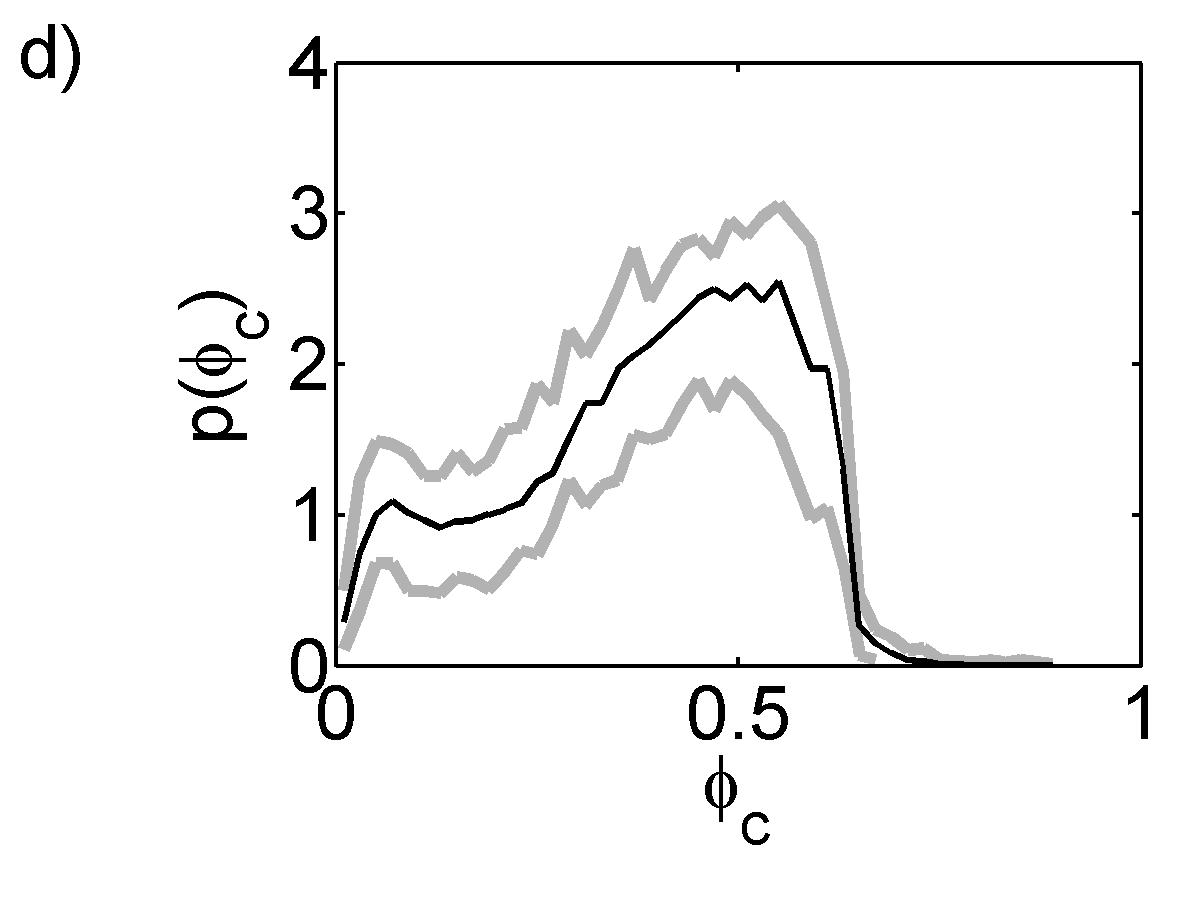}} \\
\end{center}
\caption{Relative frequency distributions of the topological cell degrees $k_c$ (a), geometric cell degrees $\kappa_c$ (b), cell areas $A_c$ (c), and form factors $\phi_c$ (d). The black line represents the mean taken over all cities, and grey lines indicate the minimum and maximum frequencies obtained in the individual networks for given $k_c$, $\kappa_c$, $A_c$ and $\phi_c$, respectively.}
\label{fig:cell}
\end{figure}

\subsection{Cell degree distributions}

In contrast to the nodes, the possible degrees of the cells (in both topological and geometric sense) are not restricted to very low numbers. Figure~\ref{fig:cell}a shows that among the 20 German cities, there is again a common behaviour of the topological cell degree distribution $p(k_c)$ with a fat tail for $k_c\geq 4$. Similar observations can be made for the associated geometric cell degree $\kappa_c$ ($\kappa_c\geq k_c$, see Fig.~\ref{fig:cell}b). The average topological cell degrees are of the order of 5, with only rather weak variations between values of 4.95 (Karlsruhe) and 5.34 (Bielefeld) as shown in Tab.~\ref{tab:GermanChar12b}. The meaning of the topological cell degree can also be illustrated by considering an adjoint graph representation based on these two-dimensional objects, i.e., a network where the nodes represent the cells of the physical road networks and the links stand for common boundaries between the cells. It follows from the above results that such adjoint graphs form approximately scale-free networks, since the topological cell degree $k_c$ in the physical space corresponds to the node degree $k_n^*$ in the associated adjoint representation of the cells.

\subsection{Form factor distributions}

Concerning the form factors, we again find very similar distributions for all cities (Fig.~\ref{fig:cell}d), with preferred values between about 0.3 and 0.6. These numbers are in good agreement with the maximum possible values for cells with degree $k_c=3$ and 4,
\begin{eqnarray}
\phi_c^{max}(k_c=3)&=&\frac{3\sqrt{3}}{4\pi} \approx 0.41, \\
\phi_c^{max}(k_c=4)&=&\frac{2}{\pi} \approx 0.64,
\end{eqnarray}
\noindent
where the latter one is the most abundant type (Fig.~\ref{fig:cell}a). Among the different cities, the average form factor of all cells varies (see Tab.~\ref{tab:GermanChar12b}) in a range between 0.352 (Bremen) and 0.411 (Dresden and Mannheim), and the standard deviation of $\phi_c$ for the individual networks takes values between 0.159 and 0.175.

\begin{figure*}[t!]
	\centering
\resizebox{\columnwidth}{!}{\includegraphics{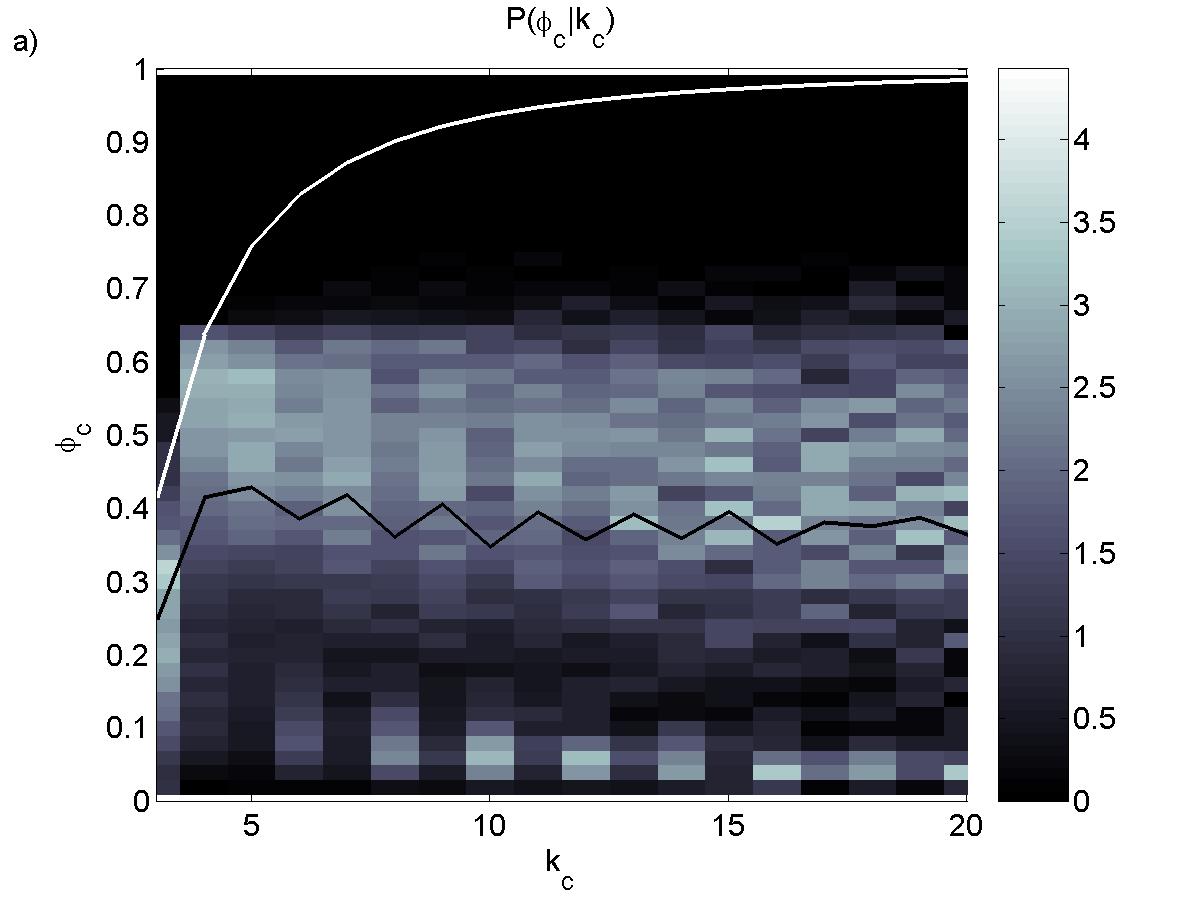}} \hfill
\resizebox{\columnwidth}{!}{\includegraphics{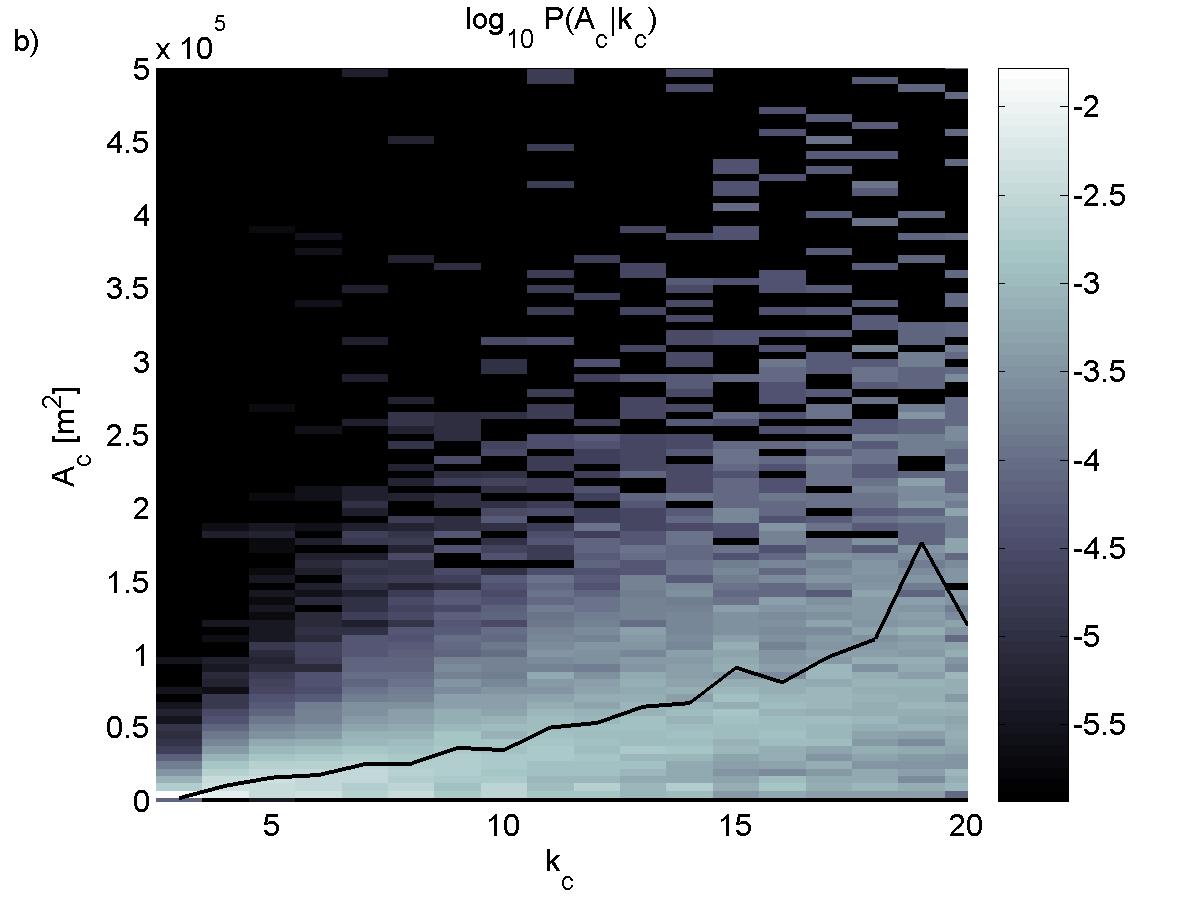}} \\
\resizebox{\columnwidth}{!}{\includegraphics{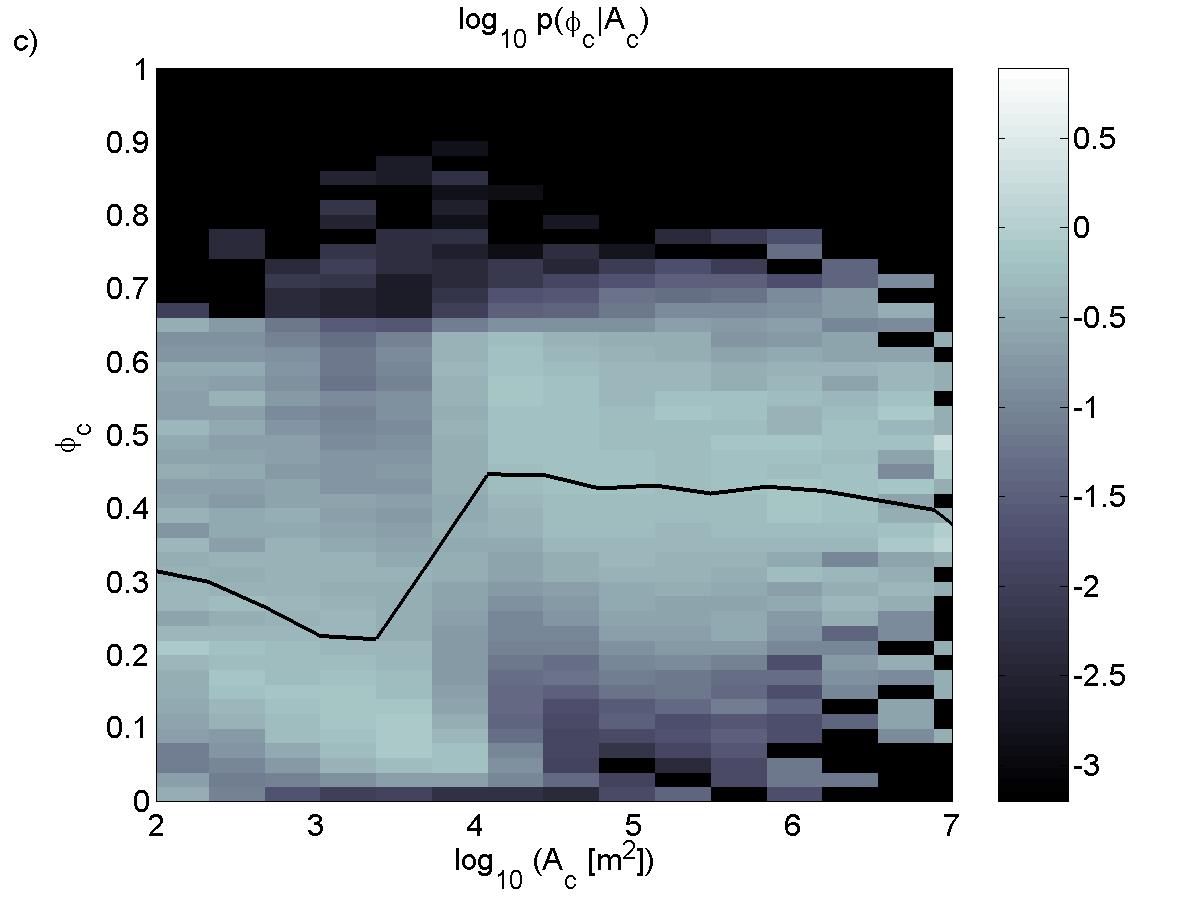}} \hfill
\resizebox{\columnwidth}{!}{\includegraphics{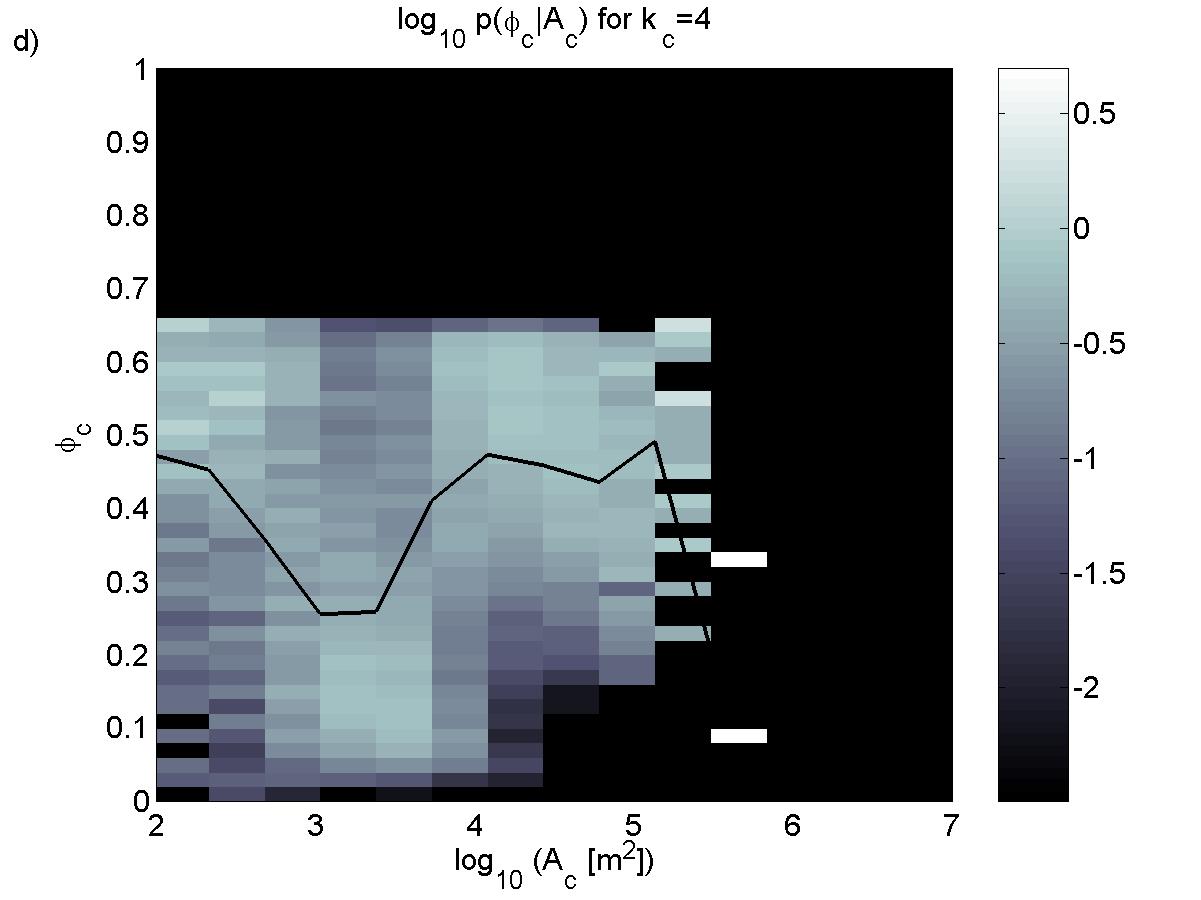}} \\
\caption{a,b) Frequency distributions (colour-coded) of form factors $\phi_c$ (a) and cell areas $A_c$ (b, in logarithmic scale) in dependence on the topological cell degrees $k_c$. The white solid line indicates the theoretical maximum value according to Eq.~(\ref{phimax}), while the black lines correspond to the mean values as a function of the cell degree. c,d) Frequency distributions (colour-coded) of the form factors in dependence on the cell area (in logarithmic scale) for all cells (c) and for cells with $k_c=4$ only (d). Black lines again indicate mean form factors in dependence on the cell area.}
\label{fig7}
\end{figure*}

Considering the interrelationships between the different cell characteristics, we expect a dependence of the form factor $\phi_c$ on the cell degree $k_c$. In particular, $\phi_c$ is bounded from above by the maximum area of a cell formed by a given number of nodes. For a prescribed perimeter $p_c$, this maximum is taken in the case of a regular $k_c$-polygon. Thus, we find
\begin{equation}
\phi_c^{max}(k_c)=\frac{k_c}{\pi}\sin\left(\frac{k_c-2}{2k_c}\pi\right)\sin\left(\frac{1}{k_c}\pi\right)
\label{phimax}
\end{equation}
\noindent
with $\phi_c^{max}\to 1$ for large $k_c$. For the German cities, there is, however, hardly any cell with $\phi_c>0.64$ (i.e., the value of $\phi_c^{max}$ for $k_c=4$). This finding underlines that even cells with a high degree mainly have rectangular shape, which is again a consequence of the preferred form of buildings and specific functional areas within urban settlements also reflected in the distributions of link and double-angles. Even more, Fig.~\ref{fig7}a shows that the most frequent form factors are even smaller than $0.64$, which suggests that many of these rectangular cells of high degree have elongated rather than {squarish} shapes. 

In order to further support the latter interpretation, the dependence of the form factor distribution $p(\phi_c)$ on the topological cell degrees has been explicitly studied. From the results shown in Fig.\ \ref{fig7}a, we identify an increasing probability of very low form factors for high cell degrees, i.e., cells with a possibly more complex (less rectangular) shape. Moreover, there is hardly any cell with a form factor above 0.64 even for $k_c\gg 4$. Both observations suggest that among the cells with large degree, there is a significant subset characterised by strongly elongated shapes. Such cells may be especially observed along geographical boundaries (e.g., rivers, highways, or railways) or in specific functional areas (e.g., large parks or industrial areas). Our aforementioned findings are supplemented by rather weak negative correlations between the average topological cell degrees and form factors of the 20 considered road networks ($r=-0.31$, $\rho=-0.31$).

\subsection{Cell area distributions}

Like for all other quantities studied so far, the distributions of cell areas also show a common behaviour for all considered German cities (see Fig.~\ref{fig:cell}c). In particular, the qualitative characteristics are similar to those of the link lengths, with almost constant cell size probabilities below about $10^4$ m$^2$ and a fat tail for larger cell areas. However, the behaviour at small sizes is distinctively different, i.e., small cells are observed more frequently than those of intermediate size, whereas in the link length distributions, a plateau-like saturation behaviour has been found (see Sec.~\ref{sec:links}). One reason for this is the influence of the cell shape: there are many cells that are exclusively formed by short road segments, some arising from a combination of short and long links (i.e., with elongated structures), and only a few cells exclusively resulting from very long links. In the following, the corresponding interrelationships will be discussed in some detail.

First, we note that for a known link length distribution, the cell areas are closely related to the cell degrees, the corresponding relationship being mediated through the associated form factors. For the respective mean values computed for the 20 individual cities, we actually find that the relationship between cell areas and cell degrees is characterised by a linear correlation coefficient of $r=0.76$ ($\rho=0.67$), which is remarkably high. 

In order to better understand the dependence of the cell area on the cell degree, let us assume that all links forming the network would have the same length $L$ (of course, this is only possible for rather generic network structures such as triangular, {square}, or hexagonal grids). In such a case, the maximum possible cell area (which is occupied by a regular $k_c$-polygon) is a function of the cell degree $k_c$ as
\begin{equation}
A_{max}(k_c)=\frac{L^2}{4}k_c\frac{1}{\tan\left(\frac{1}{k_c}\pi\right)}
\label{amax}
\end{equation}
\noindent
and approaches asymptotically the corresponding expression for the area of the circumscribed circle,
\begin{equation}
A_{max}(k_c) \to \frac{L^2}{4}\frac{\pi}{\sin^2\left(\frac{1}{k_c}\pi\right)} \approx \frac{L^2}{4} \frac{k_c^2}{\pi}.
\end{equation}

From the latter expression, it follows that the relationship between cell areas and link lengths is at most quadratic. However, in real-world urban road networks, such a dependence cannot be found. On the one hand, even simple rectangular cells of degree $k_c=4$ show a non-trivial distribution of the aspect ratio $\Delta_c$ (which is related to the form factor as $\phi_c=4\pi^{-1}\Delta_c/(\Delta_c^2+1)$). In particular, because the assumption of constant link lengths is violated, systematic deviations from the quadratic dependence arise. On the other hand, since even cells of higher degree typically have rectangular rather than higher-order polygonial shape, the typical cell areas scale clearly sub-quadratically with $k_c$ (Fig.~\ref{fig7}b). An alternative explanation for this observation could be that cells with a large degree have a tendency towards being formed by links with shorter lengths. However, the average values for the 20 German cities indicate that there is in contrast a significant positive correlation between link lengths and cell degrees ($r=0.70$, $\rho=0.52$) as well as cell areas ($r=0.93$, $\rho=0.83$).

Finally, we observe a non-trivial dependence of the form factors on the corresponding cell areas, which is shown in Fig.~\ref{fig7}c. In particular, it turns out that cells covering a large area have a tendency towards more {squarish} shapes (i.e., larger form factors), while for smaller cells ($A_c\lesssim 10^4$ m$^2$), the average form factors decreases significantly, which implies more elongated structures. This interpretation is supported when considering only the subset of cells with a topological cell degree of $k_c=4$ (Fig.~\ref{fig7}d). However, from the latter subset, we find that very small cells ($A_c\lesssim 10^3$ m$^2$) again show a clear tendency towards {squarish} shapes (i.e., higher $\phi_c$). These findings show that the typical aspect ratio of cells in urban road networks is related with the cell size in a complicated way. Following our previous considerations, we note that the empirically observed dependence is most probably determined by both building sizes and transport efficiency. In particular, we argue that the transport efficiency decreases for small cells (due to frequent stops of traffic) as well as for large cells (increase of travel distance due to the predominant rectangular structures). In this respect, the emergence of more elongated shapes for cells of intermediate size could be regarded as a certain trade-off, including parallel roads with a small spacing and perpendicular roads with larger mutual distances as a typical spatial pattern. The fact that small cells {approach again less elongated shapes} is probably related to the fact that these are often located in certain functional areas (e.g., residential areas) with a rather low traffic volume, which would imply that the dependence of cell shapes on cell sizes is at least partly influenced by the expected traffic conditions.

\subsection{Radial distributions of cell properties}

In addition to the aforementioned relationships between cell and link properties on the one hand, and link and node properties (cf. Sec.~\ref{sec:links}) on the other hand, we note that it is possible to also evaluate correlations between the node degrees and the areas and degrees of the surrounding cells. From the average values of the 20 German cities, we find $r=-0.53$ ($\rho=-0.44$) and $r=-0.83$ ($\rho=-0.86$), respectively. The last result is a consequence of the fact that cells of large degree have most often rectangular shapes and are formed by sequences of links that connect nodes with degrees of 3 and 4.

\begin{figure}[t!]
\begin{center}
\resizebox{0.48\columnwidth}{!}{\includegraphics{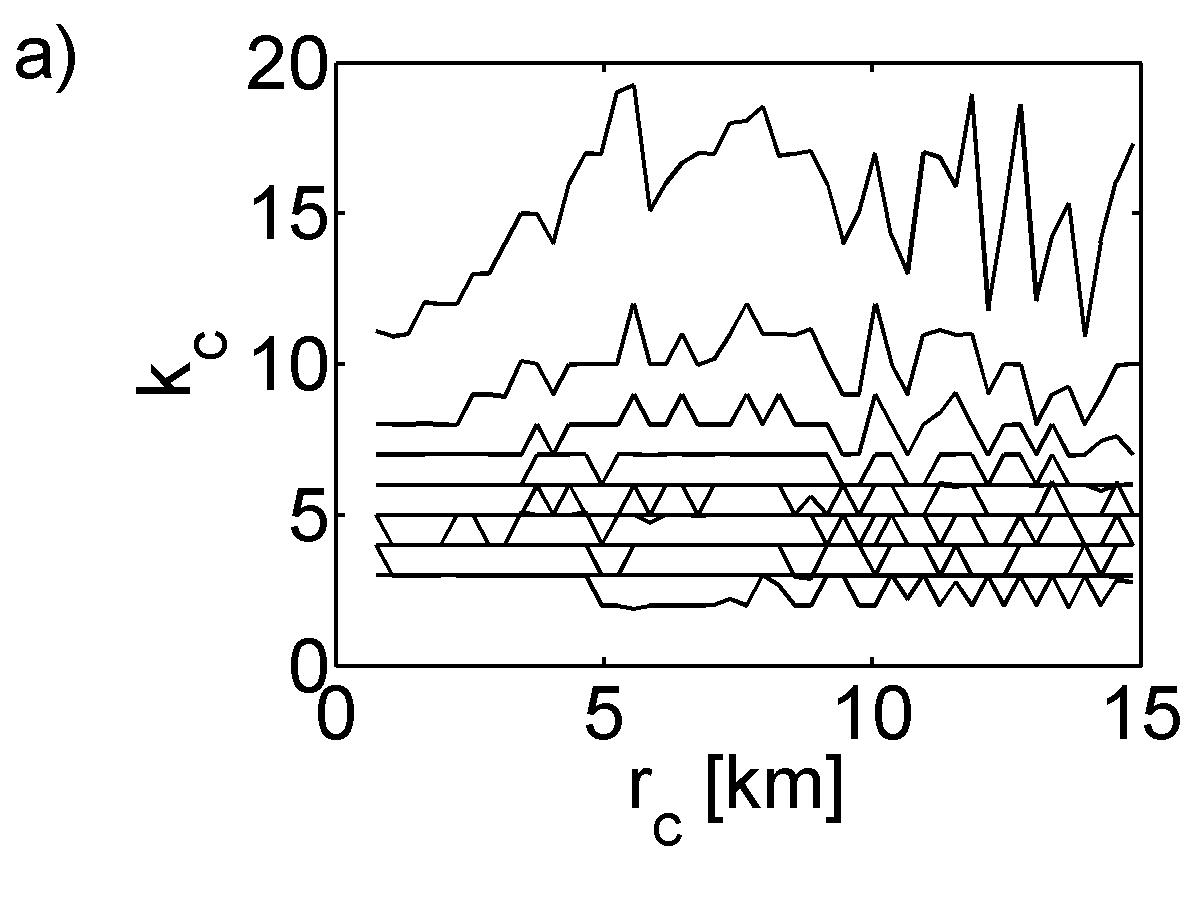}} \hfill
\resizebox{0.48\columnwidth}{!}{\includegraphics{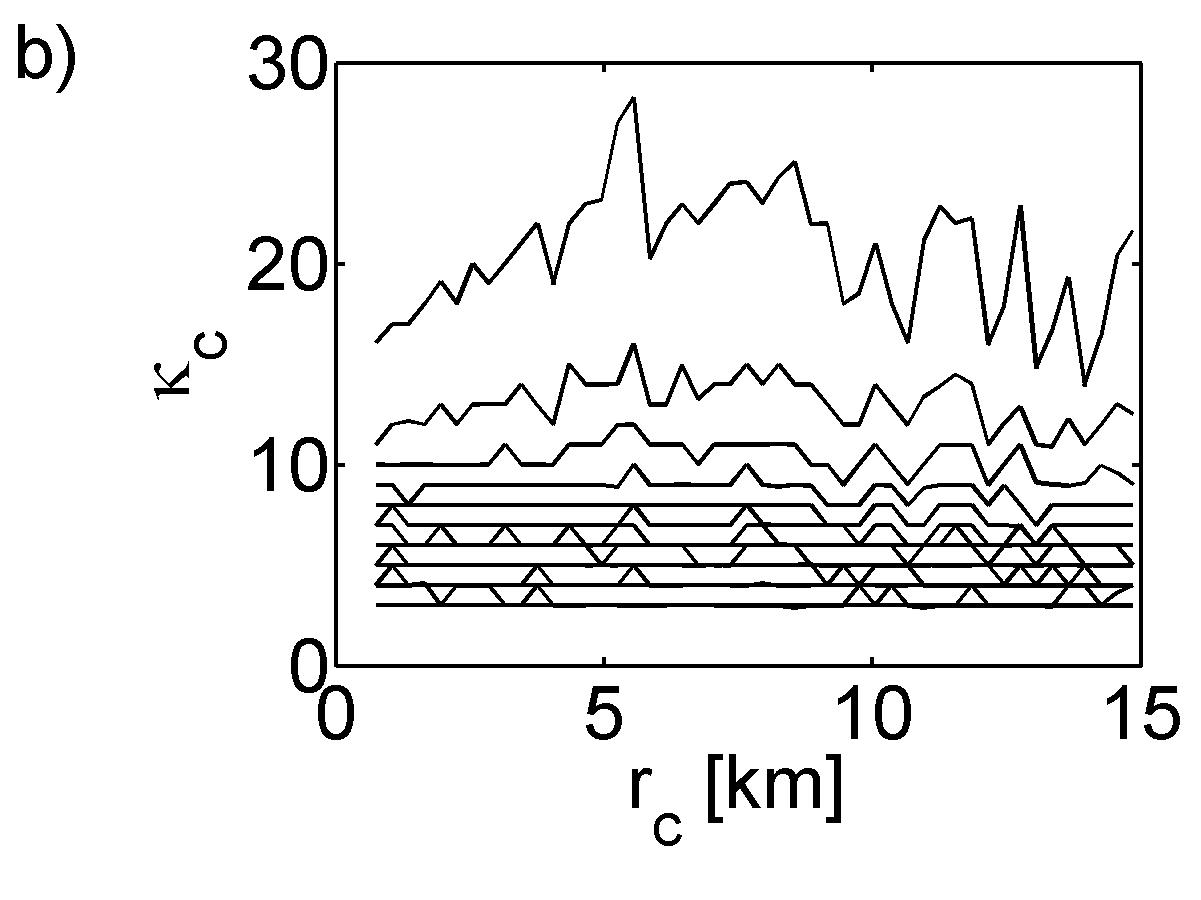}} \\
\resizebox{0.48\columnwidth}{!}{\includegraphics{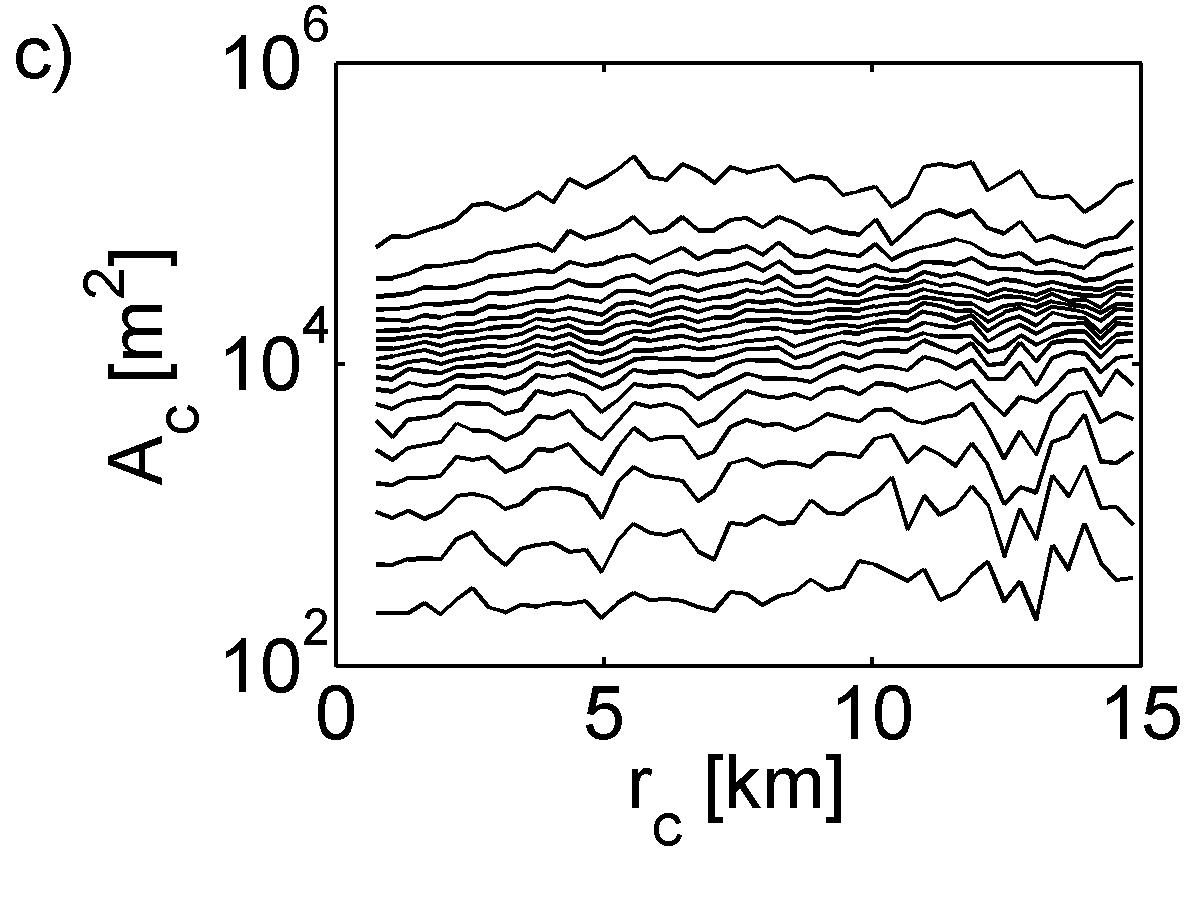}} \hfill
\resizebox{0.48\columnwidth}{!}{\includegraphics{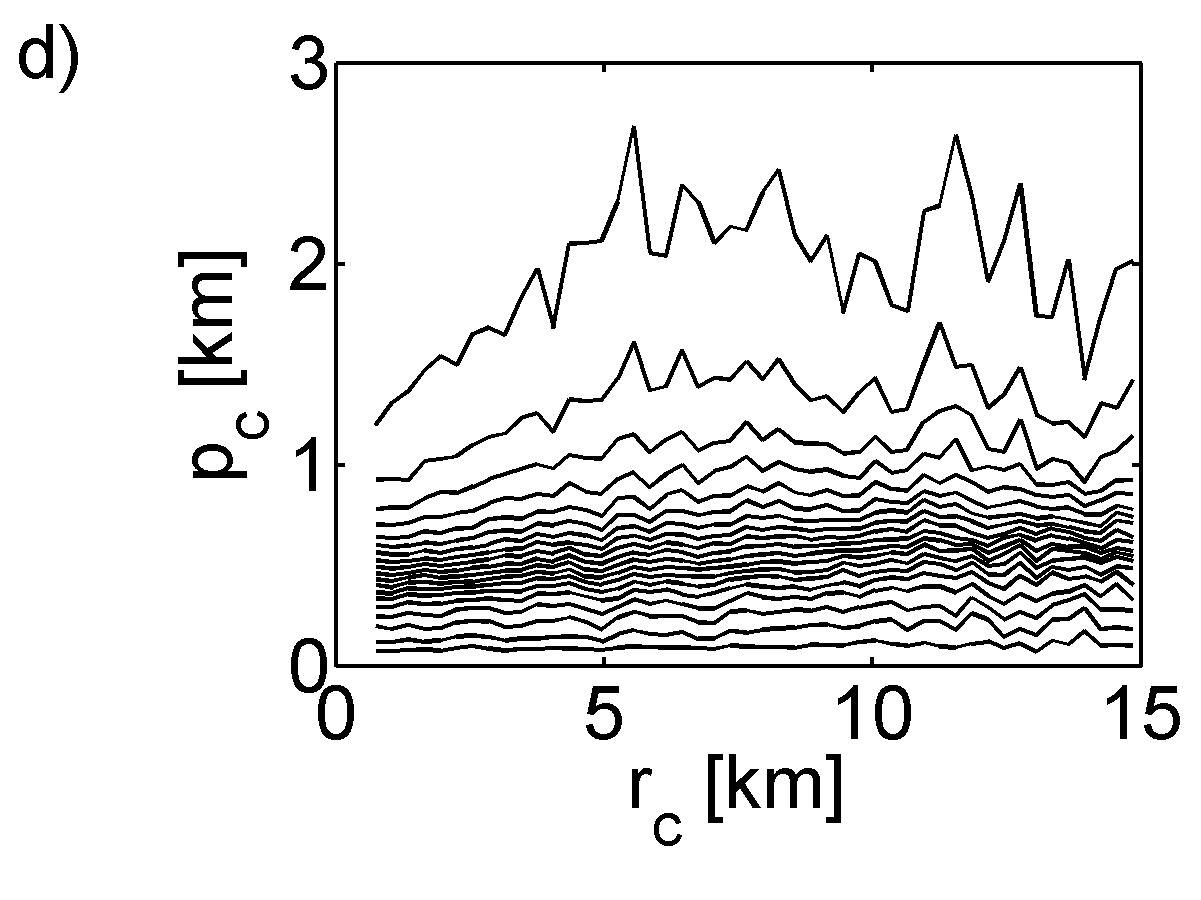}} \\
\end{center}
\caption{95\% to 5\% (in steps of 5\% from top to bottom) probability quantiles of topological cell degree $k_c$ (a), geometric cell degree $\kappa_c$ (b), cell area $A_c$ (c), and cell perimeter $p_c$ (d) obtained from the 20 considered cities in dependence of the distance of the centroid midpoints $r_c$ from the city centres.}
\label{fig:cellradial}
\end{figure}

Since both node and link properties depend on the spatial distance from the city centres (cf. Sec.~\ref{sec:links}), we expect a corresponding dependence for the cell properties as well. Masucci \textit{et~al.} \cite{Masucci2009} recently found that for the London street network, both average values and standard deviations of cell degree and cell area increase towards the outer parts of the city. Qualitatively similar findings have been reported for the distributions of areas and perimeters of individual buildings \cite{Batty2008} (which are closely related to those of the cells of the associated urban road networks). The aforementioned results coincide with our observations made for the 20 German cities, which can be seen in Fig.~\ref{fig:cellradial}. However, in contrast to the London case, we observe (as for the link lengths) a saturation of cell degrees, areas, and perimeters at intermediate distances from the city centres of about 5-10 km. This difference could be related to a possibly different structure of the analysed data as well as actual differences in the urban patterns.

As a final remark, we emphasise that the spatial heterogeneity of cell properties is related to the non-trivial large-scale morphology of the networks (Fig.~\ref{fig:example}), which is reflected in the differences of global characteristics like the population density or the spatial density of intersections $\rho_n$ (see Tab.~\ref{tab:GermanChar12}), the latter of which is directly linked with network properties like mean link length or mean cell area. Among the studied networks, we particularly find as the two extreme cases that (i) Bielefeld has many large cells of high degrees, but a low average node degree (i.e., many $k_n=3$ nodes), whereas (ii) Mannheim has much fewer large cells and, thus, a lower mean cell degree. The different relative frequencies of large cells most probably indicate the presence of distinct functional areas (industry, business, parks, etc.) or geographic constraints (mountains, lakes, or rivers, etc.).

\section{Discussion}

Recent investigations on the properties of road networks in physical space have often been restricted to node properties (i.e., distributions of degrees and other centrality measures) discussed for small parts of urban areas only \cite{Buhl2006,Scellato2006,Cardillo2006,Crucitti2006a,Crucitti2006b,Porta2006a,Porta2006b}. These studies revealed distinct differences between the network properties of historically grown (self-organised) and planned cities, as discussed in Sec.~3. In contrast, we have studied various \textit{geometric} properties of the complete urban road networks of 20 German cities, which strongly differ among each other in terms of their historical development and large-scale morphology. Nevertheless, a large degree of universality is found in all considered properties, which reflects common design principles in both historic and modern cities such as maximum area usage, optimisation of traffic conditions (with respect to necessary stops and turning processes), and minimum construction and maintenance costs. The combination of the aforementioned principles implies the presence of perpendicular structures \cite{Gastner2006} as supported by our empirical findings.

In the last years, studies on the network properties of urban road systems in physical space have been supplemented by a number of investigations addressing the topological features of the corresponding connectivity graphs. For example, Jiang and Calamund \cite{Jiang2004a} found that urban street networks can be classified as small-world networks with small average path length and high degree of clustering. Kalapala \textit{et~al.} \cite{Kalapala2006} reported a universal scaling behaviour $p(k_n)\sim k_n^{-\gamma}$ in the degree distribution $p(k_n)$ obtained from the dual representations of the national road networks of the United States, England, and Denmark, with exponents $\gamma$ of 2.2 to 2.4. This result is in certain contradiction to findings of Jiang and Calamund \cite{Jiang2004a} and Rosvall \textit{et~al.} \cite{Rosvall2005}, who did not observe power-laws in the connectivity distributions obtained from small-scale urban road networks. Wagner \cite{Wagner2008} found that the distance distribution in the dual graphs is described by a rescaled Poissonian, the parameters of which are closely correlated with the scaling of the power-law distribution of the rank-ordered street lengths.

The contradicting results on the emergence of scaling in the connectivity distributions of the dual graphs can again (at least in some parts) be understood by a distinction between self-organised and planned cities \cite{Cardillo2006}. Porta \textit{et~al.} \cite{Porta2006b} showed that a power-law with an exponent of $\gamma=2.5\pm 0.1$ describes the degree distribution of the road network of the naturally grown city of Ahmedabad, whereas other networks studied in their work showed either no scaling at all or included too few intersections to conclude with a sufficient significance about the presence of a power law. Jiang \cite{Jiang2007} found a distinct scaling in a set of 46 urban road networks with scaling exponents $\gamma$ between 1.7 and 2.6. In addition, he reported that in planned cities, very long streets (i.e., nodes of the dual graph with a high connectivity) result in a second scaling region with additional exponents of about 3.4 to 4.2. 

From the aforementioned results, we conclude that the fact that historical urban structures have grown over much longer time intervals makes their statistical features being rather specific. The set of German cities studied in this contribution can still be considered as a relatively homogeneous sample in the sense that the geographical conditions of the study area, the total age of the cities, and the economic and demographic history determining the evolution of the urban areas are {sufficiently similar}. This is {also} reflected by the universality of the geometric properties reported in this paper. In turn, this argument implies that the presented results do not provide enough empirical arguments for an emergence of comparable statistical properties in cities with completely different historical or cultural background. It will therefore be an important question for future research to which extent the presented results can be generalised independent of the region or history of a city. Moreover, for the German cities studied in this work, we can hardly conclude whether or not historical and modern parts have comparable geometric properties, since there is no clear distinction between both types of areas due to multiple small historical centres (e.g., in the case of Berlin), past damages of historical areas due to wars and governmental policies, and other related effects.

From a conceptual point of view, the structure of naturally grown urban systems has been proposed being the result of some generic self-organisation processes similar to those observed in biological systems. For example, Schaur \cite{Schaur1991} presented a comparison of some simple network properties of human settlements with those of biological systems such as dragonfly wings or maple leaves. Buhl \textit{et~al.} \cite{Buhl2004} performed extensive experimental studies on the development of ant galleries in restricted planar areas. They found a variety of different network patterns with average node degrees $\left<k_n\right>$ between 2.0 and 3.06 and single-scale exponential degree distributions. Similar results have recently been reported by Perna \textit{et~al.} \cite{Perna2008} for three-di\-men\-sio\-nal gallery networks of termite nests (with $\left<k_n\right>$ between 2.06 and 3.27). Due to our preprocessing for reducing the road networks to their geographically relevant structures, we are not able to systematically compare the results for the German cities with those obtained for other (biological as well as technological) transportation networks. A corresponding analysis will be subject of future studies.

{We emphasise that in the framework of the presented purely geometric analysis, we are not able to definitely distinguish between small-scale structures resulting from either historical (self-organised) growth or careful urban planning. Since both factors can be expected to apply in our case study, our findings suggest that some fundamental local geometric properties of urban road networks are indeed almost independent of the particular history of a city. While the relevant spatial dimensions (i.e., link lengths and cell areas) of the basic units show a substantial scatter between the studied road networks, the corresponding distribution functions are -- qualitatively -- extremely similar. Therefore, we speculate that the physical sizes of basic units can show some dependence on the historical evolution of the network (e.g., smaller cells and shorter links in historical city centres as revealed by our analysis). In a similar way, specific geographic constraints and other secondary factors matter here, whereas fundamental principles related with cost and transportation efficiency apply to both historically grown as well as strategically planned cities. Accordingly, the identified ``universal'' geometric properties do not vary significantly among the road networks under study, although the particular history of the corresponding urban structures may be quite different.}

Despite the limitations discussed above, the large degree of similarity in the geometric properties of urban road networks found in this study {clearly} supports the assumption of some universal design principles beyond the evolution of urban structures, which seem to apply to both historically grown as well as planned road networks. Since road capacities have not been considered in our investigations, one could assume a certain independence of structural properties from the traffic volume based on these first principles. However, we have observed certain non-trivial interrelationships between different geometric properties (reflected by the complex individual large-scale morphologies of different cities), which suggest that traffic conditions actually matter in the growth of urban structures. Hence, in the spirit of recent results regarding the self-organisation of urban settlements, we argue that there is a co-evolution of structure and dynamics (even beyond the transportation capacity of roads) already on the geometric level, which is supported by daily life experience. This implies that urban road networks have to be considered as adaptive co-evolutionary networks \cite{Gross2008,Gross2009}, the geometric properties of which arise from both self-organisation principles and planning strategies. We conclude that there is a considerable importance of first principles such as minimum travel times and maximum area usage in the evolution of urban structures, which lead to some universal structural properties of the resulting networks mainly determined by the planarity constraint \cite{Masucci2009}. These general results, however, need to be further validated by simulation studies of corresponding sophisticated mathematical models of growing networks.

{One particularly interesting recent result on possible relationships between road network geometry and traffic dynamics has been given by L\"ammer \textit{et~al.}~\cite{Laemmer2006a}, where the authors reported estimates of effective travel time-based dimensions of $d>2$ for Germany's largest cities. We note that this finding cannot be directly related with the outcome of our geometric analysis, since the definition of the dimension considered in \cite{Laemmer2006a} has been based on traffic dynamics instead of geometric network structures. Conceptually, we suggest that higher ``dynamic dimensions'' could be related to a strong heterogeneity in the distribution of travel times, which can be triggered by bottlenecks such as few bridges connecting different parts of a city (that have high internal connectivity), i.e., the presence of well-expressed modular structures in the network. In this spirit, the dynamic dimensions of \cite{Laemmer2006a} are related to meso-scale network properties rather than local geometric features of urban road systems.}

\section{Outlook}

We have demonstrated that urban road networks are a class of transportation systems with distinct geometric properties (in geographical space), which are controlled by planarity as well as cost and transportation efficiency. The observed similarity of the distributions of various geometric quantities among a set of cities with heterogeneous large-scale morphology suggests that some of these features, in particular, the predominance of rectangular structures and the emergence of fat tails in object-size distributions, result from universal construction principles, which calls for additional modelling studies on this phenomenon. In turn, recent results indicated that at least the presence of fat tails in the cell area distributions could arise as a natural consequence of the planarity of urban road networks \cite{Masucci2009}, since similar properties can be recovered for rather general planar network models. Hence, it will be an important question for further (both analytical as well as modelling) studies to distinguish between properties directly resulting from planarity constraints and those that are originated in dynamically relevant self-organisation mechanisms beyond the network growth. In this respect, there is also a need for deeper studies on potential relationships between the emergence of allometric scaling, which is related to power-laws in the rank-order distributions of geometric objects \cite{Batty2008}, and the fat tails of physical object-size distributions observed in both empirical data and simple models of urban growth \cite{Laemmer2006a,Masucci2009,Barthelemy2008a,Barthelemy2008b}. In particular, it has to be studied in some detail whether or not the corresponding tails actually correspond to power-laws with a universal scaling exponent of $2$ for (at least a certain class of) planar networks, as it has been recently suggested \cite{Masucci2009}, and to which extent this is reflected by the corresponding link length distributions.

As a second main line of future research, we emphasise the need of a spatially resolved analysis which allows distinguishing between different construction times, functional areas, etc. A corresponding detailed study has been beyond the scope of this comparative investigation on different road networks, but is needed in order to better distinguish between the two modes (``natural'' and ``planned'') of historical city growth. In particular, the considered sample of cities needs to be systematically extended in order to gain substantial information about the influences of regional, topographic, and functional aspects on the properties of urban road networks. Corresponding studies will probably benefit from an additional consideration of the statistical properties of the associated dual graphs, which have been suggested to allow better distinguishing between historical and planned urban areas.

{Finally, related to our observation of rather different large-scale morphologies of urban road systems, it will be relevant to study also network properties on larger scales, such as their modular structure, and relate the corresponding results to our findings presented in this work. One promising approach is utilising other measures of centrality than the node degree, such as node and link betweenness, which provide alternative characterisations of the importance of individual nodes and links in the network. We emphasise that unlike for the simple geometric properties studied here, it is likely that the degree of similarity (or even universality) in the frequency distributions of such higher-order network properties is considerably lower. A detailed corresponding investigation will be subject of future work.}

\begin{acknowledgement}
The authors wish to thank the German Research Foundation (DFG research projects He 2789/5-1,7-1,8-1,8-2), the Gottlieb Daimler und Karl Benz foundation (project ``BioLogistics''), the Volkswagen foundation (project no. I/82697), the Max Planck Society, and the Leibniz society (project ``ECONS'') for partial financial support of this work. Helpful comments of Dirk Helbing and Johannes H\"ofener are gratefully acknowledged.
\end{acknowledgement}


\begin{thebibliography}{99}

\bibitem{Strogatz2001} S.~H. Strogatz, Nature \textbf{410}, 268 (2001)
\bibitem{Albert2002} R. Albert and A.-L. Barab\'asi, Rev.\ Mod.\ Phys.\ \textbf{74}, 47 (2002)
\bibitem{Dorogovtsev2002} S.N. Dorogovtsev and J.F.F. Mendes, Adv.\ Phys.\ \textbf{51}, 1079 (2002)
\bibitem{Newman2003} M.~E.~J. Newman, SIAM Rev.\ \textbf{45}, 167 (2003)
\bibitem{Boccaletti2006} S. Boccaletti, V. Latora, Y. Moreno, M. Chavez, and D.-U. Hwang, Phys.\ Rep.\ \textbf{426}, 175 (2006)
\bibitem{Watts1999} D.~J. Watts, \textit{Small Worlds: The Dynamics of Networks between Order and Randomness} (Princeton University Press, Princeton, 1999)
\bibitem{Bornholdt2003} S. Bornholdt and H.~G. Schuster (eds.), \textit{Handbook of Graphs and Networks: From the Genome to the Internet} (Wiley-VCH, Weinheim, 2003)
\bibitem{Dorogovtsev2003} S.~N. Dorogovtesev and J.~F.~F. Mendes, \textit{Evolution of Networks} (Oxford University Press, Oxford, 2003)
\bibitem{Pastor-Satorras2003} R. Pastor-Satorras, M. Rubi, and A. Diaz-Guilera (eds.), \textit{Statistical Mechanics of Complex Networks} (Springer, Berlin, 2003)
\bibitem{Ben-Naim2004} E. Ben-Naim, H. Frauenfelder, and Z. Toroczkai (eds.), \textit{Complex Networks} (Springer, Berlin, 2004)

\bibitem{Marchiori2000} M. Marchiori and V. Latora, Physica A \textbf{285}, 539 (2000)
\bibitem{Latora2001} V. Latora and M. Marchiori, Phys.\ Rev.\ Lett.\ \textbf{87}, 198701 (2001)
\bibitem{Latora2002} V. Latora and M. Marchiori, Physica A \textbf{314}, 109 (2002)
\bibitem{Sen2003} P. Sen, S. Dasgupta, A. Chatterjee, P.~A. Sreeram, G. Mukherjee, and S.~S. Manna, Phys.\ Rev.\ E \textbf{67}, 036106 (2003)
\bibitem{Seaton2004} K.~A. Seaton and L.~M. Hackett, Physica A \textbf{339}, 635 (2004)
\bibitem{Vragovic2005} I. Vragovi\'c, E. Louis, and A. D\'{\i}az-Guilera, Phys.\ Rev.\ E \textbf{71}, 036122 (2005)
\bibitem{Kurant2006a} M. Kurant and P. Thiran, Phys.\ Rev.\ Lett.\ \textbf{96}, 138701 (2006)
\bibitem{Kurant2006b} M. Kurant and P. Thiran, Phys.\ Rev.\ E \textbf{74}, 036114 (2006)
\bibitem{Chang2006} K.~H. Chang, K. Kim, H. Oshima, and S.-M. Yoon, J.\ Korean Phys.\ Soc.\ \textbf{48}, S143 (2006)
\bibitem{Xu2007} Z. Xu and D.~Z. Sui, J.\ Geograph.\ Syst.\ \textbf{9}, 189 (2007)
\bibitem{Li2007} W. Li and X. Cai, Physica A \textbf{382}, 693 (2007)
\bibitem{Lee2008} K. Lee, W.-S. Jung, J.~S. Park, and M.~Y. Choi, Physica A \textbf{387}, 6231 (2008)
\bibitem{Ru2008} W. Ru, T. Jiang-Xia, W. Xin, W. Du-Juan, and C. Xu, Physica A \textbf{387}, 5639 (2008)
\bibitem{Domenech2009} A. Dom\'enech, Physica A \textbf{388}, 4658 (2009)

\bibitem{Amaral2000} L.~A.~N. Amaral, A. Scala, M. Barth\'el\'emy, and H.~E. Stanley, Proc.\ Natl.\ Acad.\ Sci.\ USA \textbf{97}, 11149 (2000)
\bibitem{Chi2003} L.-P. Chi, R. Wang, H. Su, X.-P. Xu, J.-S. Zhao, W. Li, and X. Cai, Chin.\ Phys.\ Lett.\ \textbf{20}, 1393 (2003)
\bibitem{Li2004} W. Li and X. Cai, Phys.\ Rev.\ E \textbf{69}, 046106 (2004)
\bibitem{Barrat2004} A. Barrat, M. Barth\'el\'emy, R. Pastor-Satorras, and A. Vespignani, Proc.\ Natl.\ Acad.\ Sci.\ USA \textbf{101}, 3747 (2004)
\bibitem{Guimera2004} R. Guimer\`a and L.~A.~N. Amaral, Europ.\ Phys.\ J.\ B \textbf{38}, 381 (2004)
\bibitem{Guimera2005} R. Guimer\`a, S. Mossa, A. Turtschi, and L.~A.~N. Amaral, Proc.\ Natl.\ Acad.\ Sci.\ USA \textbf{102}, 7794 (2005)
\bibitem{Wang2005} R. Wang and X. Cai, Chin.\ Phys.\ Lett.\ \textbf{22}, 2715 (2005)
\bibitem{Li2006b} W. Li, Q.~A. Wang, L. Nivanen, and A. Le M\'ehaut\'e, Physica A \textbf{368}, 262 (2006)
\bibitem{Guida2007} M. Guida and F. Maria, Chaos Solitons Fractals \textbf{31}, 527 (2007)
\bibitem{Bagler2008} G. Bagler, Physica A \textbf{387}, 2972 (2008)
\bibitem{Bagler2009} G. Bagler, \textit{Complex Network view of performance and rosks on Airport Networks}, in \textit{Airports: Performance, Risks, and Problems}, edited by P.~B. Larauge and M.~E. Castille (Nova, Hauppauge, 2009), pp. 199--205
\bibitem{Correa2008} L.~E. Correa da Rocha, J.\ Stat.\ Mech.\ Theory Exper., P04020 (2009)
\bibitem{Wu2004} J. Wu, Z. Gao, H. Sun, and H. Huang, Mod.\ Phys.\ Lett.\ B \textbf{18}, 1043 (2004)
\bibitem{Sienkiewicz2005} J. Sienkiewicz and J.~A. Ho{\l}yst, Phys.\ Rev.\ E \textbf{72}, 046127 (2005)
\bibitem{Sienkiewicz2005b} J. Sienkiewicz and J.~A. Ho{\l}yst, Acta Phys.\ Polon.\ b \textbf{36}, 1771 (2005)
\bibitem{Gastner2006b} M.~T. Gastner and M.~E.~J. Newman, J.\ Stat.\ Mech.\ Theory Exper., P01015 (2006)
\bibitem{Li2006} P. Li, X. Xiong, Z.-L. Qiao, G.-Q. Yuan, X. Sun, and B.-H. Wang, Chin.\ Phys.\ Lett.\ \textbf{23}, 3384 (2006)
\bibitem{vonFerber2007} C. von Ferber, T. Holovatch, Yu. Holovatch, and V. Palchykov, Physica A \textbf{380}, 585 (2007)
\bibitem{vonFerber2009} C. von Ferber, T. Holovatch, Yu. Holovatch, and V. Palchykov, Eur.\ Phys.\ J.\ B \textbf{68}, 261 (2009)
\bibitem{Berche2009} B. Berche, C. von Ferber, T. Holovatch, and Yu. Holovatch, Eur.\ Phys.\ J.\ B \textbf{71}, 125 (2009)
\bibitem{Xu2007b} X. Xu, J. Hu, and F. Liu, Chaos \textbf{17}, 023129 (2007)
\bibitem{Hu2008} Y. Hu and D. Zhu, Physica A \textbf{388}, 2061 (2009)
\bibitem{Kaluza2010} P. Kaluza, A. K\"olzsch, M.~T. Gastner, and B. Blasius, J.\ R.\ Soc.\ Interface {\bf 7}, 1093 (2010)
\bibitem{Gastner2006} M.~T. Gastner and M.~E.~J. Newman, Eur.\ Phys.\ J.\ B \textbf{49}, 247 (2006)

\bibitem{Erdoes1959} P. Erd{\H{o}}s and A. R\'enyi, Publ.\ Math.\ Debrecen \textbf{6}, 290 (1959)
\bibitem{Bollobas2001} B. Bollob\'as, \textit{Random Graphs} (2nd ed., Cambridge University Press, Cambridge, 2001)
\bibitem{Barabasi1999} A.-L. Barab\'asi and R. Albert, Science \textbf{286}, 509 (1999)
\bibitem{Dorogovtsev2000} S.~N. Dorogovtsev, J.~F.~F. Mendes, and A.~N. Samukhin, Phys.\ Rev.\ Lett.\ \textbf{85}, 4633 (2000)
\bibitem{Caldarelli2007} G. Caldarelli, \textit{Scale-Free Networks - Complex Webs in Nature and Technology} (Oxford University Press, Oxford, 2007)

\bibitem{Schaur1991} E. Schaur, \textit{Ungeplante Siedlungen / Non-planned Settlements} (Kr\"amer, Stuttgart, 1991)
\bibitem{Frankhauser1994} P. Franckhauser, \textit{La Fractalit\'e des Structures urbaines} (Anthropos, Paris, 1994)
\bibitem{Schweitzer1997} F. Schweitzer (ed.), \textit{Self-Organization of Complex Structures: From Individual to Complex Dynamics} (Gordon and Breach, London, 1997)
\bibitem{Weidlich1999} W. Weidlich and G. Haag, \textit{An Integrated Model of Transport and Urban Evolution: With an Application to a Metropole of an Emerging Nation} (Springer, Berlin, 1999)
\bibitem{Weidlich2000} W. Weidlich, \textit{Sociodynamics: A Systematic Approach to Mathematical Modelling in the Social Sciences} (Harwood Academic Publishers, New York, 2000)
\bibitem{Humpert2002} K. Humpert, K. Brenner, and S. Becker (eds.), \textit{Fundamental Principles of Urban Growth} (M\"uller and Busmann, Dortmund, 2002)
\bibitem{Schweitzer2003} F. Schweitzer, \textit{Brownian Agents and Active Particles. Collective Dynamics in the Natural and Social Sciences} (Springer, Berlin, 2003)
\bibitem{Batty2005} M. Batty, \textit{Cities and Complexity. Understanding Cities with Cellular Automata, Agent-Based Models, and Fractals} (MIT Press, Cambridge, 2005)
\bibitem{Kuehnert2006} C. K\"uhnert, D. Helbing, and G.~B. West, Physica A {\bf 363}, 96 (2006)
\bibitem{Bettencourt2007} L.~M.~A. Bettencourt, J. Lobo, D. Helbing, C. K\"uhnert, and G.~B. West, Proc. Natl. Acad. Sci. USA {\bf 104}, 7301 (2007)
\bibitem{Helbing2009} D. Helbing, C. K\"uhnert, S. L\"ammer, A. Johansson, B. Gehlsen, H. Ammoser, and G.~B. West, \textit{Power laws in urban supply networks, social systems, and dense pedestrian crowds}, in \textit{Complexity Perspectives in innovation and Social Change}, edited by D. Lane, S. van der Leeuw, D. Pumain, and G.~B. West (Springer, Berlin, 2009), pp. 433-450

\bibitem{Jiang2004a} B. Jiang and C. Claramunt, Environ.\ Plan.\ B: Plan.\ Des.\ {\bf 31}, 151 (2004)
\bibitem{Jiang2004b} B. Jiang and C. Claramunt, GeoInformatica {\bf 8}, 157 (2004)
\bibitem{Laemmer2006a} S. L\"ammer, B. Gehlsen, and D. Helbing, Physica A \textbf{363}, 89 (2006)
\bibitem{Buhl2006} J. Buhl, J. Gautrais, N. Reeves, R.~V. Sol\'e, S. Valverde, P. Kuntz, and G. Theraulaz, Eur.\ Phys.\ J.\ B \textbf{49}, 513 (2006)
\bibitem{Scellato2006} S. Scellato, A. Cardillo, V. Latora, and S. Porta, Eur.\ Phys.\ J.\ B \textbf{50}, 221 (2006)
\bibitem{Cardillo2006} A. Cardillo, S. Scellato, V. Latora, and S. Porta, Phys.\ Rev.\ E \textbf{73}, 066107 (2006)
\bibitem{Crucitti2006a} P. Crucitti, V. Latora, and S. Porta, Chaos \textbf{16}, 015113 (2006)
\bibitem{Crucitti2006b} P. Crucitti, V. Latora, and S. Porta, Phys.\ Rev.\ E \textbf{73}, 036125 (2006)
\bibitem{Porta2006a} S. Porta, P. Crucitti, and V. Latora, Environ.\ Plan.\ B: Plan.\ Des.\ \textbf{33}, 705 (2006)
\bibitem{Porta2006b} S. Porta, P. Crucitti, and V. Latora, Physica A \textbf{369}, 853 (2006)
\bibitem{Kalapala2006} V. Kalapala, V. Sanwalani, A. Clauset, and C. Moore, Phys.\ Rev.\ E {\bf 73}, 026130 (2006)

\bibitem{West1996} G.~B. West, \textit{Introduction to Graph Theory} (Prentice Hall, Upper Saddle River, 1996)
\bibitem{Masucci2009} A.~P. Masucci, D. Smith, A. Crooks, and M. Batty, Eur.\ Phys.\ J.\ B {\bf 71}, 259 (2009)
\bibitem{Laherrere1998} J. Laherr\`ere and D. Sornette, Eur.\ Phys.\ J.\ B \textbf{2}, 525 (1998)
\bibitem{Roehner1998} B.~M. Roehner and D. Sornette, Eur.\ Phys.\ J.\ B \textbf{4}, 387 (1998)
\bibitem{McCauley2003} J.~L. McCauley and G.~H. Gunaratne, Physica A \textbf{329}, 178 (2003)
\bibitem{Mallat1989} S.~G. Mallat, IEEE Trans.\ Pat.\ Anal.\ Mach.\ Intell.\ \textbf{11}, 674 (1989)
\bibitem{Everitt1981} B.~S. Everitt and D.~J. Hand, \textit{Finite Mixture Distributions} (Chapmann and Hall, London, 1981)
\bibitem{Godreche1992} C. Godr\`eche, I. Kostov, and I. Yekutieli, Phys.\ Rev.\ Lett.\ {\bf 69}, 2674 (1992)
\bibitem{Barthelemy2008a} M. Barth\'elemy and A. Flammini, Phys.\ Rev.\ Lett.\ \textbf{100}, 138702 (2008)
\bibitem{Barthelemy2008b} M. Barth\'elemy and A. Flammini, Netw.\ Spat.\ Econ.\ \textbf{9}, 401 (2009)
\bibitem{Batty2008} M. Batty, R. Carvalho, A. Hudson-Smith, R. Milton, D. Smith, and P. Steadman, Eur.\ Phys.\ J.\ B \textbf{63}, 303 (2008)

\bibitem{Rosvall2005} M. Rosvall, A. Trusina, P. Minnhagen, and K. Sneppen, Phys.\ Rev.\ Lett.\ {\bf 94}, 028701 (2005)
\bibitem{Wagner2008} R. Wagner, Physica A \textbf{387}, 2120 (2008)
\bibitem{Jiang2007} B. Jiang, Physica A \textbf{384}, 647 (2007)
\bibitem{Buhl2004} J. Buhl, J. Gautrais, R.~V. Sol\'e, P. Kuntz, S. Valverde, J.~L. Deneubourg, and G. Theraulaz, Eur.\ Phys.\ J.\ B \textbf{42}, 123 (2004)
\bibitem{Perna2008} A. Perna, S. Valverde, J. Gautrais, C. Jost, R. Sol\'e, P. Kuntz, and G. Theraulaz, Physica A \textbf{387}, 6235 (2008)
\bibitem{Gross2008} T. Gross and B. Blasius, J.\ Roy.\ Soc.\ Interface {\bf 5}, 259 (2008)
\bibitem{Gross2009} T. Gross and H. Sayama (eds.), \textit{Adaptive Networks -- Theory, Models and Applications} (Springer, Berlin, 2009)

\end{thebibliography}
\end{document}